\documentclass[10pt]{article} 
\usepackage[preprint]{tmlr}

\usepackage{amsmath,amsfonts,bm}

\def\eqref#1{equation~\ref{#1}}

\def\1{\bm{1}}

\DeclareMathAlphabet{\mathsfit}{\encodingdefault}{\sfdefault}{m}{sl}
\SetMathAlphabet{\mathsfit}{bold}{\encodingdefault}{\sfdefault}{bx}{n}

\usepackage{pdfpages}

\usepackage{url}
\usepackage{wrapfig}
\usepackage{amsmath,amsfonts}
\usepackage{algorithmic}
\usepackage{algorithm}
\usepackage{array}
\usepackage[caption=false,font=normalsize,labelfont=sf,textfont=sf]{subfig}
\usepackage{textcomp}
\usepackage{stfloats}
\usepackage{hyperref}
\usepackage{tcolorbox}
\usepackage{graphicx} 
\usepackage{url}
\usepackage{verbatim}
\newcommand{\tablogo}[1]{\makebox[1.2cm][c]{\raisebox{-0.1cm}{\includegraphics[width=1.2cm, height=0.35cm, keepaspectratio]{#1}}}}
\usepackage{enumitem}
\usepackage{pdfpages}
\usepackage{placeins}

\usepackage{adjustbox}
\usepackage{pifont}
\usepackage{marvosym}

\newcolumntype{L}[1]{>{\raggedright\arraybackslash}p{#1}}
\newcolumntype{C}[1]{>{\centering\arraybackslash}p{#1}}

\newcommand{\YearRow}[1]{%
    \addlinespace[0.45em]
    \rowcolor{gray!12}
    \multicolumn{\TableCols}{@{}l@{}}{\hspace{0.4em}\textbf{\textit{#1}}} \\
    \addlinespace[0.15em]
}

\usepackage[utf8]{inputenc}
\usepackage{booktabs} 
\usepackage{multirow} 
\usepackage{amssymb}  
\usepackage{pifont}   
\usepackage{graphicx} 
\newcommand{\cmark}{\checkmark}
\newcommand{\xmark}{\ding{55}}
\usepackage[table]{xcolor}

\usepackage{longtable}
\usepackage{afterpage}

\title{A Survey of Large Audio Language Models: Generalization, Trustworthiness, and Outlook}

\author{
\name
Kaiwen Luo$^{1,*}$, Zhenhong Zhou$^{1,*}$, Leyan Wang$^{2,*}$,
Liang Lin$^{1,*}$, Tianyu Shao$^{3}$, Yuanhe Zhang$^{4}$,
Yang Xiao$^{5}$, Yuxuan Li$^{6}$, Miao Yu$^{7}$, Kailin Lyu$^{8}$,
Jiaming Zhang$^{1}$, Li Sun$^{4}$, Songze Li$^{9}$, Yueming Wu$^{10}$,
Ting Dang$^{5}$, Xiaojun Jia$^{1}$, Dongrui Liu$^{11}$, Kai Li$^{12}$,
Rohan Kumar Das$^{13}$, Siyuan Liang$^{1}$, Xinfeng Li$^{1}$,
Qiankun Li$^{1}$, Jing Chen$^{14}$, Xingjun Ma$^{15}$,
Kun Wang$^{1,\text{\Letter}}$, Junhao Dong$^{1,\text{\Letter}}$,
Deqing Zou$^{10}$, Yu Cheng$^{16}$, Xia Hu$^{11}$, Zhigang Zeng$^{10}$,
Sen Su$^{17}$, Yang Liu$^{1}$, Yu-Gang Jiang$^{15}$,
Philip S. Yu$^{18}$, Yew-Soon Ong$^{1}$
\addr
\vspace{5mm}
\\$^1$ Nanyang Technological University;\\
$^2$ Independent Researcher;\\
$^3$ North China Electric Power University;\\
$^4$ Beijing University of Posts and Telecommunications;\\
$^5$ The University of Melbourne;\\
$^6$ University of Chinese Academy of Sciences;\\
$^7$ University of Science and Technology of China;\\
$^8$ Institute of Automation, Chinese Academy of Sciences;\\
$^9$ Southeast University;\\
$^{10}$ Huazhong University of Science and Technology;\\
$^{11}$ Shanghai AI Laboratory;\\
$^{12}$ Tsinghua University;\\
$^{13}$ Fortemedia Singapore;\\
$^{14}$ Wuhan University;\\
$^{15}$ Fudan University;\\
$^{16}$ Chinese University of Hong Kong;\\
$^{17}$ Chongqing University of Posts and Telecommunications;\\
$^{18}$ University of Illinois Chicago\\[0.5em]
{\small\normalfont
$^*$ These authors contributed equally.\quad
$\text{\Letter}$ Corresponding authors.}
}

\begin{document}
\includepdf[
  pages=1,
  pagecommand={\thispagestyle{empty}},
  fitpaper=true
]{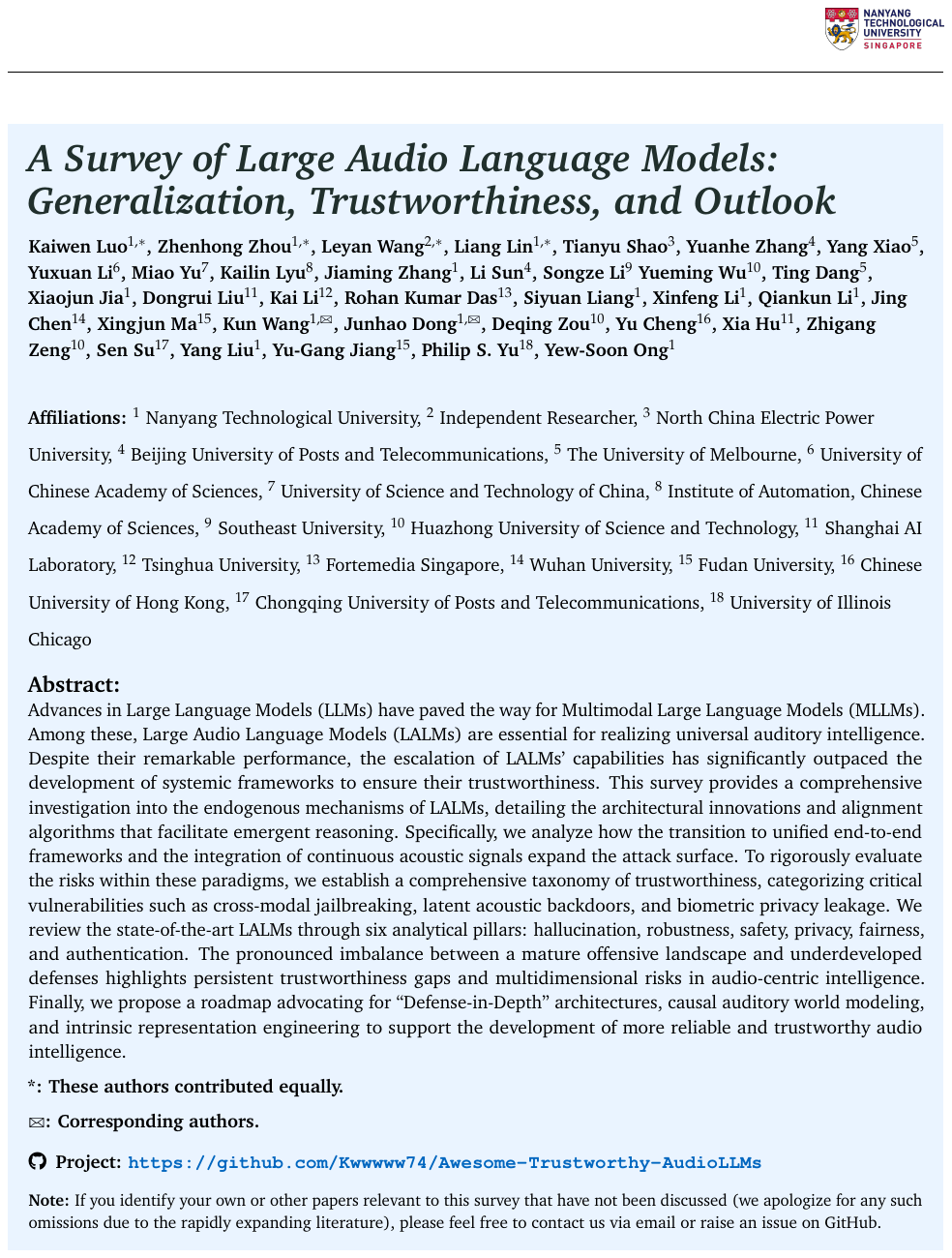}

\setcounter{page}{1}

\maketitle

\begin{abstract}
Advances in Large Language Models (LLMs) have paved the way for Multimodal Large Language Models (MLLMs). Among these, Large Audio Language Models (LALMs) are essential for realizing universal auditory intelligence. Despite their remarkable performance, the escalation of LALMs' capabilities has significantly outpaced the development of systemic frameworks to ensure their trustworthiness. This survey provides a comprehensive investigation into the endogenous mechanisms of LALMs, detailing the architectural innovations and alignment algorithms that facilitate emergent reasoning. Specifically, we analyze how the transition to unified end-to-end frameworks and the integration of continuous acoustic signals expand the attack surface. To rigorously evaluate the risks within these paradigms, we establish a comprehensive taxonomy of trustworthiness, categorizing critical vulnerabilities such as cross-modal jailbreaking, latent acoustic backdoors, and biometric privacy leakage. We review the state-of-the-art LALMs through six analytical pillars: hallucination, robustness, safety, privacy, fairness, and authentication. The pronounced imbalance between a mature offensive landscape and underdeveloped defenses highlights persistent trustworthiness gaps and multidimensional risks in audio-centric intelligence. Finally, we propose a roadmap advocating for ``Defense-in-Depth'' architectures, causal auditory world modeling, and intrinsic representation engineering to support the development of more reliable and trustworthy audio intelligence.
\end{abstract}

\section{Introduction}

The emergence of Large Language Models (LLMs) \cite {ouyang2022training, achiam2023gpt,touvron2023llama, bai2023qwen, liu2024deepseek, guo2025deepseek} has transformed the landscape of artificial intelligence, establishing a robust foundation for the transition toward unified multimodal frameworks. This evolution into Multimodal Large Language Models (MLLMs) \cite{yan2025position, yan2025survey, team2026qwen3} is designed to emulate the multi-sensory nature of human perception across diverse sensory inputs. Among human senses, \textbf{audio} represents a primary medium for human communication and perception of the environment \cite{latif2023sparks}, as it carries a vast amount of information within its signal. Previous research in audio intelligence relied on modular systems designed for a single task, such as automatic speech recognition \cite{wang2026audio, shi2026qwen3} or sound classification \cite{gemmeke2017audio, kong2020panns}. The recent transition from these artifacts to unified \textbf{Large Audio Language Models (LALMs)} \cite{chu2023qwen, tang2023salmonn, rubenstein2023audiopalm, chu2024qwen2, wu2025step} represents a step for universal audio intelligence.

Despite these remarkable advancements in auditory capabilities, the organic integration of language and audio modalities introduces complex safety and alignment challenges. Textual LLMs primarily address vulnerabilities within discrete text \cite{shi2024large, wang2025comprehensive, yu2025survey, ma2026safety}. In contrast, LALMs introduce the audio modality, which presents an intricate risk landscape \cite{lin2025hidden, chen2025synthetic, aloufi2026evaluation, chen2026hijacking} due to the continuous properties of the acoustic signal. The deployment of LALMs within critical sectors further expands this complex risk landscape, translating these continuous-signal vulnerabilities into real-world threats. However, while the development of these capabilities is expanding, the research landscape remains fragmented and lacks a unified roadmap. Existing research predominantly details architectural innovations \cite{sakshi2025spur, alex2025pal, you2026world} or specific concerns \cite{luong2025llamapartialspoof, li2025dfallm, nguyen2026analyzing}, yet there remains a significant lack of work dedicated to a systematic taxonomy of the safety implications for these systems. Recognizing that intrinsic trustworthiness cannot be guaranteed without a deep understanding of the underlying architecture, this research fragmentation highlights the necessity for a structured review that bridges the gap between mechanisms and safety.

While foundational overviews and reviews of speech models \cite{latif2023sparks, peng2025survey, su2025audio} offer comprehensive insights into auditory perception, they often treat safety and ethical considerations as peripheral topics. Similarly, recent literature focused on evaluation provides a framework  \cite{yang2025towards} for assessing model behavior but lacks a systematic taxonomy of the underlying security threats and safety mechanisms. Although an earlier review has addressed trustworthiness in speech \cite{feng2022review}, they precede the recent shift toward unified generative frameworks, focusing largely on traditional machine learning. And specialized surveys remain predominantly concentrated on singular issues such as the detection of deepfakes and biometric authentication \cite{yi2023audio, li2025survey, pham2025comprehensive}. A comparison with these existing audio surveys is provided in Table \ref{tab:1}, illustrating the lack of literature dedicated to the implications of trustworthiness of these models.

\newcommand{\TableCols}{14}
\begin{table}[htbp]
\centering
\caption{Comparison with existing surveys. }
\setlength{\tabcolsep}{0.3pt} 
\renewcommand{\arraystretch}{1.13} 

\begin{tabular*}{\columnwidth}{@{\extracolsep{\fill}}l c cccccc ccccc c @{}}
\toprule
\multirow{2}{*}{\textbf{Survey}} & \multirow{2}{*}{\textbf{Obj.$^{\ddagger}$}} & \multicolumn{6}{c}{\textbf{Trustworthiness$^{\dagger}$}} & \multicolumn{5}{c}{\textbf{Stage$^{\star}$}} & \multirow{2}{*}{\textbf{O}} \\ 
\cmidrule(lr){3-8} \cmidrule(lr){9-13}
& & \textbf{H} & \textbf{P} & \textbf{F} & \textbf{S} & \textbf{R} & \textbf{A} & \textbf{DP} & \textbf{PT} & \textbf{FT} & \textbf{DE} & \textbf{EV} & \\ 
\midrule
\YearRow{Year 2022}
\cite{feng2022review} & S & \xmark & \cmark & \cmark & \xmark & \cmark & \xmark & \cmark & \xmark & \xmark & \xmark & \cmark & \cmark \\
\midrule
\YearRow{Year 2023}
\cite{latif2023sparks} & A & \cmark & \cmark & \cmark & \cmark & \cmark & \cmark & \cmark & \cmark & \cmark & \cmark & \cmark & \cmark \\
\cite{yi2023audio} & S & \xmark & \xmark & \xmark & \xmark & \xmark & \cmark & \cmark & \xmark & \xmark & \xmark & \cmark & \cmark \\
\midrule
\YearRow{Year 2024}
\cite{li2025survey} & S & \xmark & \xmark & \xmark & \xmark & \xmark & \cmark & \cmark & \xmark & \xmark & \xmark & \cmark & \xmark \\
\cite{pham2025comprehensive} & S & \xmark & \xmark & \xmark & \xmark & \cmark & \cmark & \cmark & \xmark & \xmark & \cmark & \cmark & \cmark \\
\cite{peng2025survey} & A & \xmark & \xmark & \xmark & \xmark & \xmark & \xmark & \xmark & \cmark & \cmark & \cmark & \cmark & \cmark \\
\midrule
\YearRow{Year 2025}
\cite{su2025audio} & A & \cmark & \xmark & \cmark & \xmark & \cmark & \xmark & \cmark & \cmark & \cmark & \cmark & \cmark & \ \cmark \\
\cite{cui2025recent} & S+M & \cmark & \xmark & \cmark & \cmark & \cmark & \xmark & \xmark & \xmark & \xmark & \cmark & \cmark & \cmark \\
\cite{yang2025towards} & A+M & \cmark & \xmark & \cmark & \cmark & \cmark & \xmark & \xmark & \xmark & \xmark & \xmark & \cmark & \cmark \\
\midrule
\textbf{Ours} & A+S+M & \cmark & \cmark & \cmark & \cmark & \cmark & \cmark & \cmark & \cmark & \cmark & \cmark & \cmark & \cmark \\
\bottomrule
\end{tabular*}

\vspace{1.2mm}
\raggedright
\scriptsize
$^{\ddagger}$ \textbf{Object}: Audio-LLM (A), Speech-LM (S), Multi-modal LLM (M). \\
$^{\dagger}$ \textbf{Trustworthiness}: Hallucination (H), Privacy (P), Fairness (F), Safety (S), Robustness (R), Authentication (A). \\
$^{\star}$ \textbf{Stage}: Data Prep (DP), Pre-training (PT), Fine-tuning (FT), Deployment (DE), Evaluation (EV).
\textbf{Outlook}: Outlook(O)
\label{tab:1} 
\end{table}

Upon reviewing surveys and systematically investigating the related literature, we conclude that our survey endeavors to address several questions that existing surveys have not covered. The main contributions of this survey are summarized as follows:
\begin{itemize}[leftmargin=*]
   \item \textbf{Systematic Investigation of Endogenous Mechanisms:} We conduct a thorough examination of the internal structures within LALMs, detailing the structural improvements and alignment techniques that support the emergence of logical reasoning. This analysis provides the technical foundation required to understand the evolution toward unified models for auditory intelligence.
    \item \textbf{Comprehensive Trustworthy Review:} We establish a systematic classification of trustworthiness challenges, identifying critical vulnerabilities including cross-modal jailbreak through acoustic cues, latent acoustic backdoors, and biometric privacy leakage. Additionally, we evaluate the landscape of current leading models through the six pillars of trustworthiness, which consist of hallucination, robustness, safety, privacy, fairness, and authentication.
    \item \textbf{Identification of Imbalance and Future Framework:} Our analysis reveals a significant imbalance where offensive research has advanced significantly while defensive mechanisms remain limited and reactive. We propose a framework for future research, advocating for a shift toward layered defense architectures, causal auditory world modeling, and intrinsic representation engineering to achieve intrinsically trustworthy audio intelligence.
\end{itemize}

\begin{figure*}[t]
    \centering
    \includegraphics[width=\linewidth]{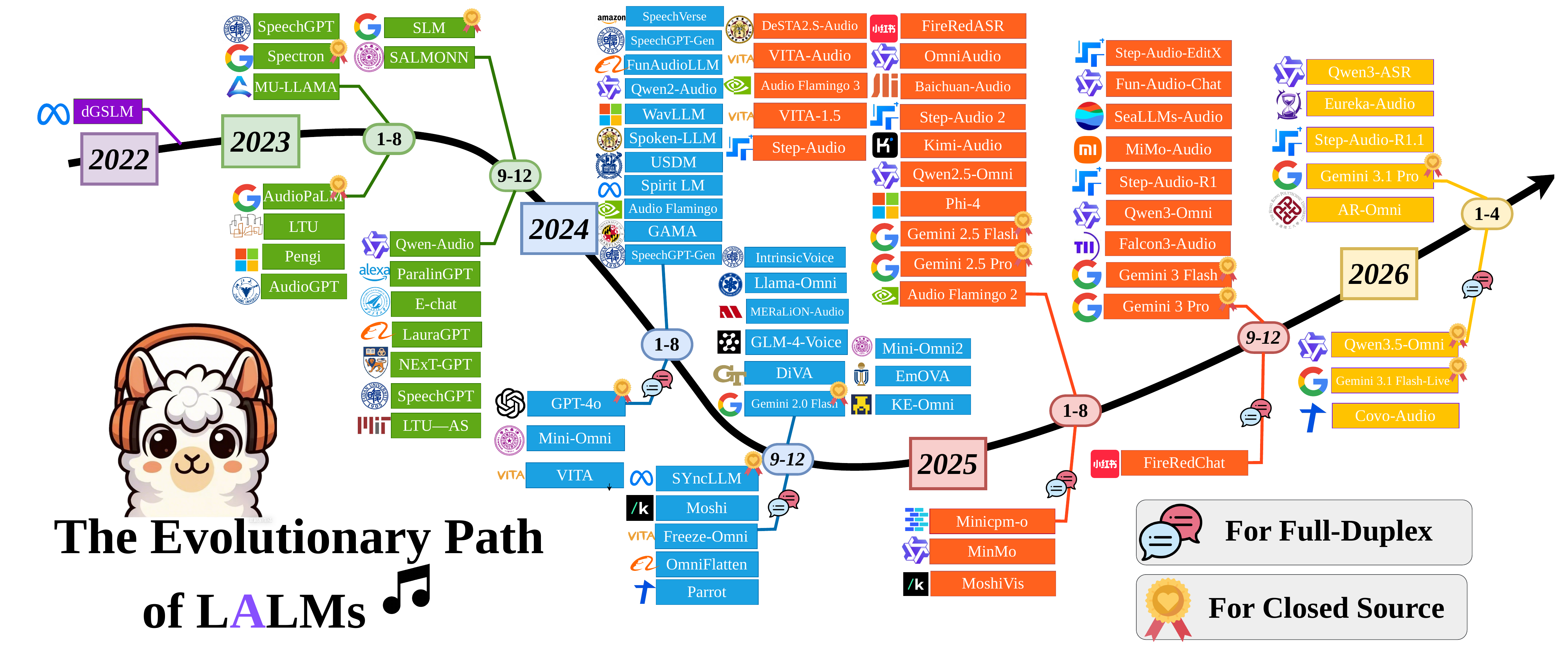}
    \caption{The Evolutionary Roadmap of LALMs from Cascaded Systems to End-to-End Causal Cognition from 2022 to 2026.}
    \label{fig:2} 
\end{figure*}

\section{Endogenous Mechanisms of LALMs}

\begin{figure*}[t]
    \centering    \includegraphics[width=\linewidth]{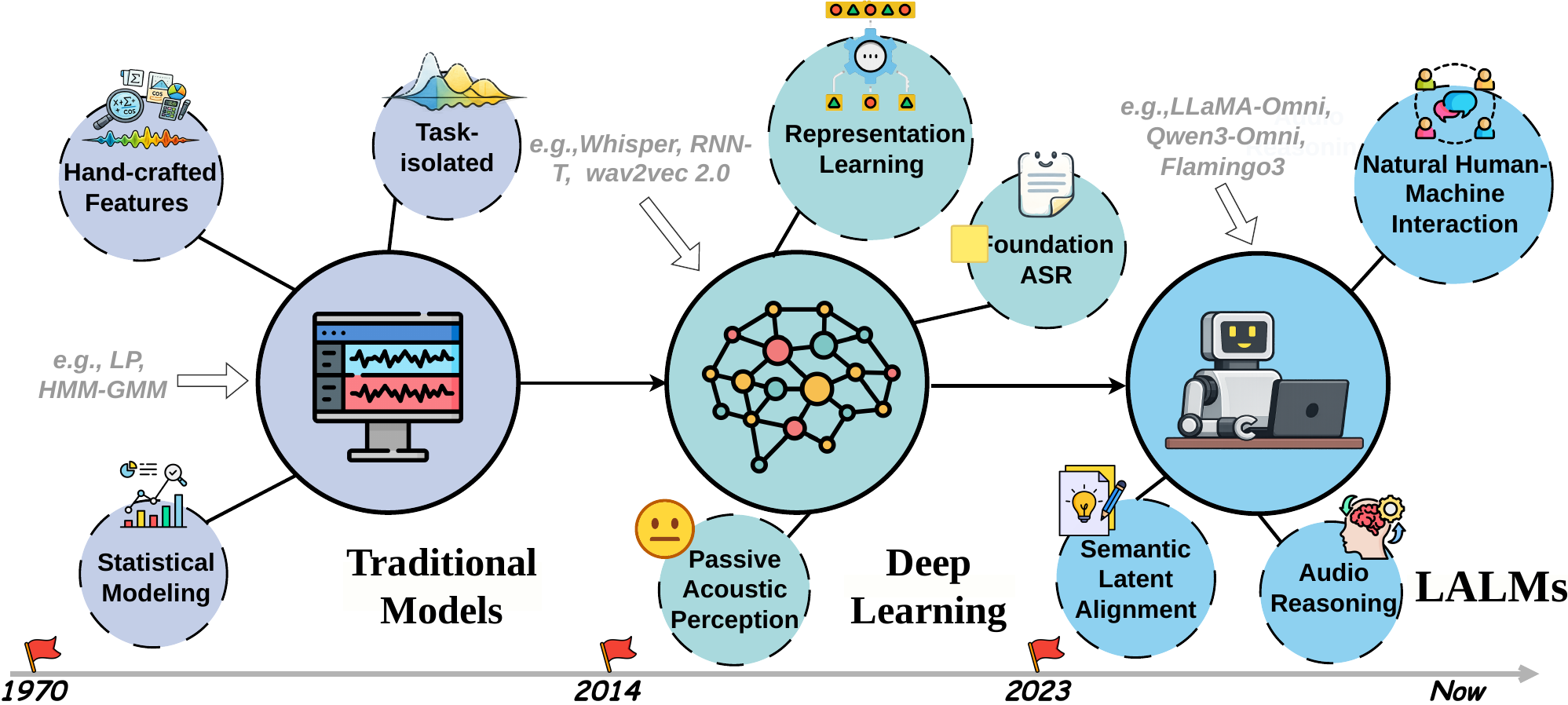}
    \caption{Architectural and Paradigmatic Evolution from Traditional Audio Models to LALMs.}
    \label{fig:1} 
\end{figure*}

This section investigates the internal mechanisms governing how LALMs process information, exploring the synergy between architectural design, representational paradigms, and optimization strategies as shown in figure \ref{fig:2} and \ref{fig:1} and table \ref{tab:2}. The fundamental capabilities of LALMs are underpinned by their architectural design and the transition from task-specific cascaded systems toward unified, end-to-end multimodal frameworks \cite{zhang2023speechgpt, rubenstein2023audiopalm}. Unlike traditional systems characterized by modular decoupling, contemporary architectures employ a sophisticated pipeline designed to map continuous, non-stationary auditory signals into structured semantic latent spaces \cite{tang2023salmonn, chu2024qwen2}.

\subsection{Architectural Foundations}
The structural integrity of LALMs is established upon a composite information processing pipeline that facilitates the translation of raw acoustic signals into semantic representations. This architectural framework  integrates three components consisting of an acoustic encoder, an alignment projector, and a LLM backbone.

The acoustic encoder functions as the foundational interface for sensory perception. Current research emphasizes the rigorous evaluation of these components through initiatives \cite{ma2026interspeech}. The investigation of information transfer mechanisms from these encoders to language decoders is essential for optimizing system performance \cite{alex2025pal}. Moreover, specialized encoding strategies are employed to characterize physical attributes including spatial descriptors \cite{jiang2026sci}.

The alignment projector and integration frameworks serve as the critical nexus between modalities. Modern architectures frequently incorporate heterogeneous sensory inputs to improve task specific precision such as the integration of visual and auditory understanding \cite{zhao2025hears}. Frameworks like \textbf{TWNM} \cite{you2026world} and \textbf{SPUR} \cite{sakshi2025spur} enhance the adaptability of existing systems. Furthermore, architectural refinements for specialized deployment scenarios are represented by egocentric multichannel processing \cite{lin2025wearvox}.

The LLM backbone provides the essential cognitive capacity for reasoning. Detailed evaluations indicate that the auditory knowledge inherently encoded within these backbones during text based pre-training significantly impacts subsequent audio grounded capabilities \cite{lu2026how}.

Significant structural innovations continue to enhance the efficiency and versatility of the LALM pipeline. These developments include the adoption of structured embeddings for integrated understanding and editing as presented in \textbf{SALM} \cite{hu2025salm}. Other paradigms propose fundamental shifts in processing methodology such as the dual-resolution parallel frameworks or the implementation of transformers that operate directly within latent spaces \cite{lu2025latent}.

\subsection{Representational Paradigms}
The selection of representational paradigms determines the efficacy and semantic grounding of LALMs. A fundamental distinction in current research involves the comparative utility of discrete audio tokens versus continuous temporal manifolds. Unified frameworks increasingly utilize text-aligned factorized audio tokenization to ensure consistency between auditory and linguistic units \cite{yang2026uniaudio}. To address the scalability constraints of discrete sequences, researchers have developed audio token compression techniques that maintain semantic density while reducing computational overhead \cite{bhati2025towards}. Additionally the capture of paralinguistic nuances is enhanced through fine-grained feature augmentation including vowel-level modifications designed to improve emotional prosody \cite{wang2026vowelprompt}. This representational choice dictates the model's trustworthiness: while discrete tokenization risks discarding critical acoustic safety cues during compression, continuous manifolds preserve rich paralinguistic nuances but consequently increase the attack possibility for adversarial vulnerabilities.

\subsection{Training and Alignment Strategies}
Architectural sparsity and parameter efficient fine tuning serve as the primary mechanisms for adapting models to complex tasks with minimal overhead. The implementation of specialized Mixture of Experts adapters addresses gradient conflicts and promotes representational disentanglement during cross modal training \cite{lei2026moe}. Efficiency is further improved through segmentwise pruning techniques that mitigate the overhead of processing continuous streams \cite{gibier2025segmentwise}. For domain specific applications, the utilization of Low Rank Adaptation facilitates precise temporal localization in high stakes environments such as clinical therapy \cite{bn2025fine}.

Systematic evaluation of these optimizations is facilitated by benchmarks \cite{luo2026chronosaudio}. Methodological advancements include the development of extended context mechanisms for long form understanding \cite{chaichana2026extending} and techniques designed to bridge temporal gaps between frames to maintain dependency capture \cite{wang2026listening}. The end-to-end contrastive pretraining models can improve performance for long form question answering capabilities \cite{hu2026end}.

Sophisticated alignment algorithms resolve modality bias in cross-modal representations. Few-shot learning enables high proficiency with minimal data \cite{zhang2025mimo}. To ensure models rely on acoustic evidence, researchers employ attention rebalancing \cite{wang2025pay} and audio contribution-aware post-training \cite{he2025measuring}. Specialized alignment strategies \cite{grinberg2026alarm} refine this synergy. Knowledge distillation transfers reasoning from vision to audio \cite{wang2025sightsound} or via weighted cross-modal distillation \cite{hu2026cord}. Advanced metrics like attention-weighted kernel alignment optimize speech emotion recognition \cite{yang2026attention}.

Complementary to training alignments, inference optimizations provide a lightweight alternative to ensure generation quality without extensive retraining. Feedback driven retrieval augmented generation improves output quality through iterative verification \cite{zhao2025feedback}. Similarly test time adaptation methods enhance robustness for emotional recognition tasks without additional training \cite{shi2025emo}.

The evolution of LALMs is defined by the transition from rigid turn-taking toward synchronous interaction \cite{ji2024wavchat, chen2025from}. This shift necessitates sophisticated architectural schemes that enable real-time conversations beyond the turn-based game \cite{zhang2024beyond}, including codec-free designs for speech understanding and generation \cite{yu2024salmonn}. While some research investigates efficient and direct duplex modeling \cite{hu2025efficient}, others highlight the potential of modular systems \cite{liu2025xtalk} or propose plug-and-play streaming state prediction modules to ensure real-time responsiveness \cite{yan2026soulx}. Optimization strategies have also advanced through time-controllable training \cite{chang2026tico}, reinforcement learning for interactivity optimization \cite{hsiao2026aspirin}, and the use of natural monologues via dual training \cite{yao2025flm}. And cognitive capabilities within duplex models are being extended through asynchronous knowledge retrieval \cite{chien2026moshirag} and latent reasoning to model internal cognition \cite{wu2026silent}. The proliferation of these full-duplex technologies has concurrently spurred the development of comprehensive evaluation frameworks to assess real-time disfluency, multi-turn dynamics, and semantic-aware interruptions. In addition, the development of privacy-preserving end-to-end dialogue models ensures secure full-duplex communication \cite{kuzmin2026privacy}, while technical reports like \textbf{Covo-Audio} \cite{wang2026covo} continue to delineate the evolving landscape of universal auditory intelligence.

\subsection{Emergent Reasoning Mechanisms}
The transition of LALMs from passive transcription engines to cognitive agents capable of complex deduction represents a pivotal advancement in auditory intelligence. This evolution is underpinned by the development of internal mechanisms that facilitate logical grounding and planning.

\begin{wrapfigure}{r}{0.48\textwidth}
    \vspace{-0.8em}
    \centering
    \includegraphics[width=0.46\textwidth]{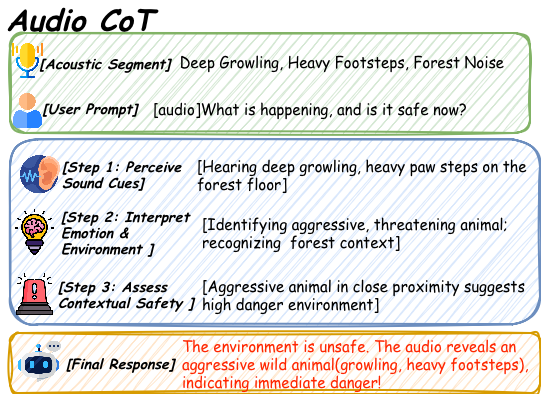}
    \caption{Visualization of a standard LALM with Audio-CoT.}
    \label{fig:audiocot}
    \vspace{-0.8em}
\end{wrapfigure}

Central to these emergent capabilities is the implementation of Audio Chain-of-Thought (Audio-CoT) architectures, compelling models to generate intermediate reasoning trajectories prior to formulating final responses \cite{xiong2025thinking} in figure \ref{fig:audiocot}. The depth of comprehension is further enhanced by embedding reasoning steps directly within the multimodal processing flow via audio-interleaved frameworks, such as \textbf{ECHO} \cite{wu2026echo}. To enable these capabilities without extensive retraining researchers have introduced training-free steering mechanisms that activate reasoning pathways by nudging the hidden states of the model \cite{ieong2026nudging}. Furthermore the necessity for real-time cognitive processing has led to investigations into whether models can maintain reasoning efficiency while simultaneously listening to continuous audio \cite{shih2025can}.

Reinforcement Learning (RL) and process oriented reward systems serve as the primary drivers for incentivizing consistent and scalable logic. By utilizing reasoning process rewards models are encouraged to maintain logical validity throughout multi-step deductions \cite{fan2025incentivizing}. This paradigm is extended to specialized domains through emotion-rule-based RL frameworks that enhance the consistency of models executing tasks in affect-rich environments as shown in \textbf{EMO-RL} \cite{li2025emo}. Other strategies employ RL to guide models on the optimal timing and methodology for initiating reasoning processes \cite{wu2025audio}. Such incentivized reasoning is critical for solving complex logical challenges and is fundamental to the advancements presented in \textbf{SoundMind} \cite{diao2025soundmind}.

Adaptability and scalability of reasoning at inference time enable models to effectively navigate ambiguity and complex high-dimensional tasks. Difficulty-adaptive mechanisms empower models to dynamically allocate computational resources based on instruction complexity \cite{sheng2025think}. To resolve highly ambiguous emotional cues, researchers utilize test-time scaling to progressively expand the computational depth of the model during decoding \cite{jia2026decoding}.

Advanced manifestations of these internal mechanisms include agentic frameworks and causal world modeling. The integration of models into agentic systems allows for multifaceted task execution and autonomous tool use \cite{wijngaard2025audiotoolagent}. Causal state-action planning pushes the boundaries of reasoning by enabling models to simulate and reason through physical world dynamics \cite{zhou2025speech} . The robustness of these reasoning capabilities is evaluated through benchmarks targeting acoustic-semantic conflicts where models must resolve contradictions between tone and lexical content \cite{huang2026tone}.

\subsection{Future Directions of LALMs' Framework}
The evolution of LALMs is moving beyond superficial pattern matching toward deep cognitive and causal intelligence. We identify four trajectories that will define the next generation of auditory reasoning engines.

First, LALMs must transition toward causal auditory world modeling, enabling counterfactual reasoning to simulate physical dynamics and event sequences within auditory scenes \cite{zhou2025speech}. Second, optimizing the efficiency-robustness Pareto frontier necessitates semantic-aware token compression \cite{bhati2025towards} and factorized tokenization \cite{yang2026uniaudio} to maintain performance across long-form contexts \cite{luo2026chronosaudio}. Third, integrating agentic frameworks with full-duplex intelligence marks the next stage of synchronous interaction \cite{ji2024wavchat}, requiring robust handling of disfluency and tool-use in real-time conversations \cite{yu2024salmonn, lin2025full}. 
Fourth, cross-modal knowledge distillation and multi-sensory alignment will empower models to ``listen between frames'' by transferring spatial reasoning from vision to audio \cite{wang2025sightsound, grinberg2026alarm}.

As these advancements expand the multimodal attack surface, the next-generation framework must pioneer intrinsic representation engineering, ensuring that emergent capabilities are grounded in trustworthiness.

\begingroup
\normalsize
\renewcommand{\TableCols}{7}
\setlength{\tabcolsep}{1.5pt}
\renewcommand{\arraystretch}{1.13}
\setlength{\LTleft}{0pt}
\setlength{\LTright}{0pt}
\setlength{\LTcapwidth}{\textwidth}

\begin{longtable}{@{\extracolsep{\fill}}
  >{\raggedright\arraybackslash}p{0.33\textwidth}
  >{\centering\arraybackslash}p{0.13\textwidth}
  >{\centering\arraybackslash}p{0.11\textwidth}
  >{\centering\arraybackslash}p{0.13\textwidth}
  >{\centering\arraybackslash}p{0.11\textwidth}
  >{\centering\arraybackslash}p{0.065\textwidth}
  >{\centering\arraybackslash}p{0.065\textwidth}
@{}}
\caption{Summary of Large Audio Language Models from 2022 to 2026.}\label{tab:2}\\
\toprule
\multirow{2}{*}{\textbf{Model}} &
\multirow{2}{*}{\textbf{Institute}} &
\multirow{2}{*}{\textbf{Release}} &
\multirow{2}{*}{\shortstack{\textbf{Base LLM}\\\textbf{Params}}} &
\multirow{2}{*}{\shortstack{\textbf{Full-}\\\textbf{Duplex}}} &
\multicolumn{2}{c}{\textbf{Multimodality}} \\
\cmidrule(lr){6-7}
& & & & & \scriptsize\textbf{Text} & \scriptsize\textbf{Audio} \\
\midrule
\endfirsthead

\multicolumn{7}{c}{%
  \tablename~\thetable{} Summary of Large Audio Language Models from 2022 to 2026 (continued).%
}\\
\toprule
\multirow{2}{*}{\textbf{Model}} &
\multirow{2}{*}{\textbf{Institute}} &
\multirow{2}{*}{\textbf{Release}} &
\multirow{2}{*}{\shortstack{\textbf{Base LLM}\\\textbf{Params}}} &
\multirow{2}{*}{\shortstack{\textbf{Full-}\\\textbf{Duplex}}} &
\multicolumn{2}{c}{\textbf{Multimodality}} \\
\cmidrule(lr){6-7}
& & & & & \scriptsize\textbf{Text} & \scriptsize\textbf{Audio} \\
\midrule
\endhead

\midrule
\multicolumn{7}{r}{\textit{Continued on next page}}\\
\endfoot

\bottomrule
\endlastfoot

\YearRow{Year 2022}
dGSLM \cite{nguyen2023generative} & \tablogo{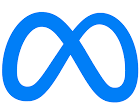} & Mar 2022 & - & \xmark & \xmark & \cmark \\

\YearRow{Year 2023}
SpeechGPT \cite{zhang2023speechgpt} & \tablogo{icon/fudan.png} & May 2023 & 13B & \xmark & \cmark & \cmark \\
Pengi \cite{deshmukh2023pengi} & \tablogo{icon/microsoft.png} & May 2023 & 124M & \xmark & \cmark & \cmark \\
LTU \cite{ramaswamy2025enhancing} & \tablogo{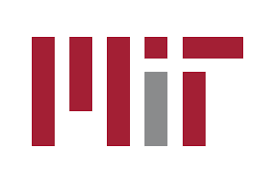} & May 2023 & 7B & \xmark & \cmark & \cmark \\
Spectron \cite{nachmani2024spoken} & \tablogo{icon/Google.jpg} & May 2023 & 350M/1B & \xmark & \cmark & \cmark \\
AudioPaLM \cite{rubenstein2023audiopalm} & \tablogo{icon/Google.jpg} & Jun 2023 & 8B & \xmark & \cmark & \cmark \\
MU-LLaMA \cite{liu2024music} & \tablogo{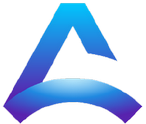} & Aug 2023 & 7B & \xmark & \cmark & \cmark \\
LTU-AS \cite{gong2023joint} & \tablogo{icon/mit.png} & Sep 2023 & 7B & \xmark & \cmark & \cmark \\
SLM \cite{wang2023slm} & \tablogo{icon/Google.jpg} & Sep 2023 & 13B & \xmark & \cmark & \cmark \\
SALMONN \cite{tang2023salmonn} & \tablogo{icon/thu.jpg} & Oct 2023 & 13B & \xmark & \cmark & \cmark \\
LauraGPT \cite{du2023lauragpt} & \tablogo{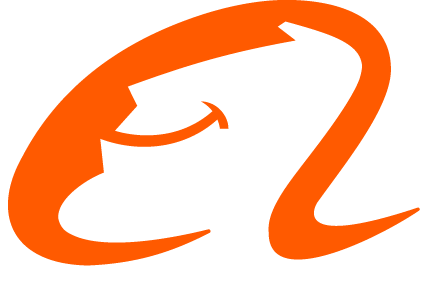} & Oct 2023 & 2B & \xmark & \cmark & \cmark \\
Qwen-Audio \cite{chu2023qwen} & \tablogo{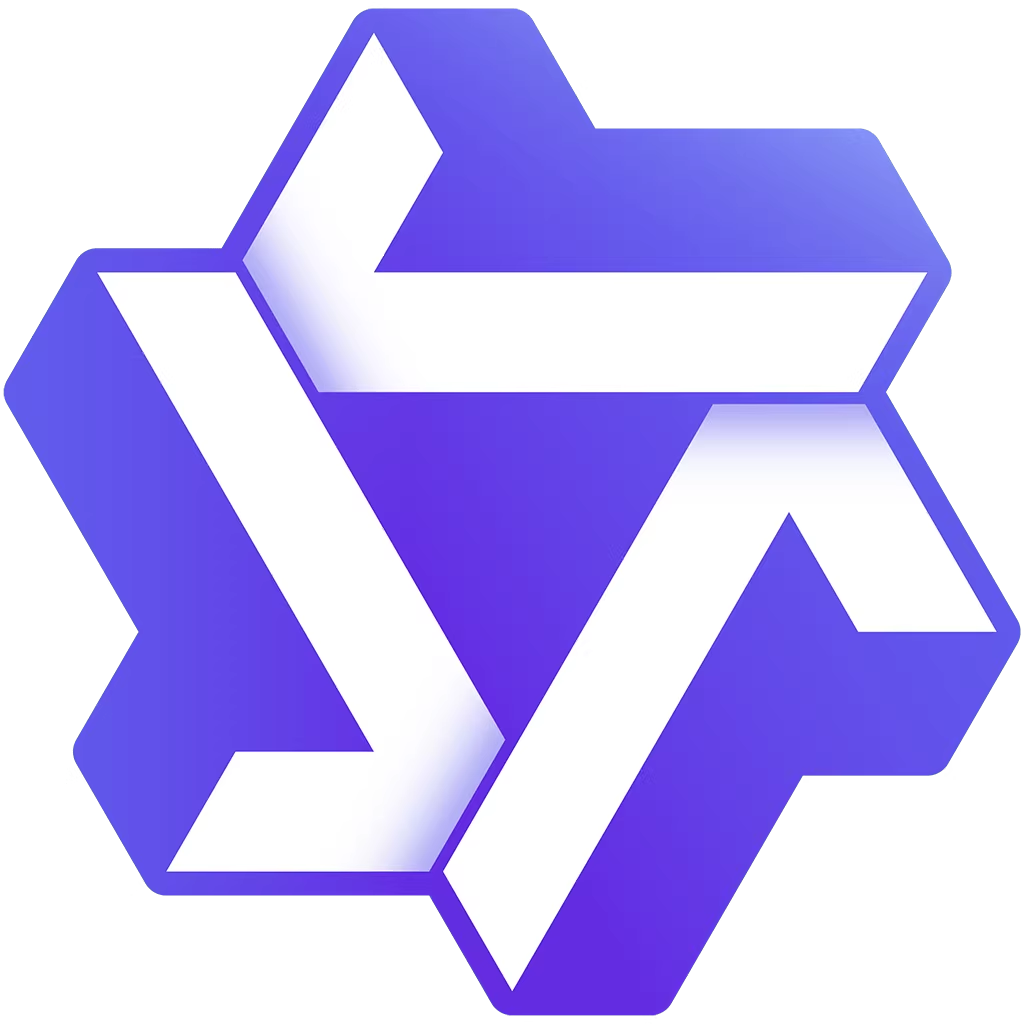} & Nov 2023 & 7B & \xmark & \cmark & \cmark \\
ParalinGPT \cite{lin2024paralinguistics} & \tablogo{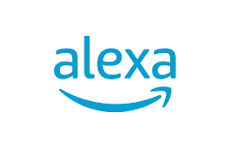} & Dec 2023 & 345M & \xmark & \cmark & \cmark \\
E-chat \cite{xue2024chat} & \tablogo{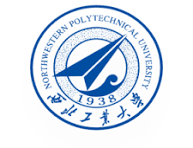} & Dec 2023 & 7B & \xmark & \cmark & \cmark \\

\YearRow{Year 2024}
SpeechGPT-Gen \cite{zhang2024speechgpt} & \tablogo{icon/fudan.png} & Jan 2024 & 7B & \xmark & \cmark & \cmark \\
Audio Flamingo \cite{kong2024audio} & \tablogo{icon/nvidia.png} & Feb 2024 & 1.3B & \xmark & \cmark & \cmark \\
Spoken-LLM \cite{lin2024advancing} & \tablogo{icon/hk.png} & Feb 2024 & 7B & \xmark & \cmark & \cmark \\
Spirit LM \cite{nguyen2025spirit} & \tablogo{icon/meta.png} & Feb 2024 & 7B & \xmark & \cmark & \cmark \\
USDM \cite{li2024emergent} & \tablogo{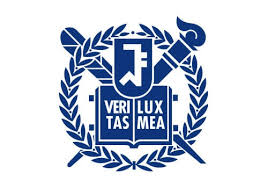} & Feb 2024 & 7B & \xmark & \cmark & \cmark \\
WavLLM \cite{hu2024wavllm} & \tablogo{icon/microsoft.png} & Mar 2024 & 7B & \xmark & \cmark & \cmark \\
SpeechVerse \cite{das2024speechverse} & \tablogo{icon/Amazon.png} & May 2024 & 3B & \xmark & \cmark & \cmark \\
GAMA \cite{ghosh2024gama} & \tablogo{icon/mu.png} & Jun 2024 & 7B & \xmark & \cmark & \cmark \\
Qwen2-Audio \cite{chu2024qwen2} & \tablogo{icon/Qwen.png} & Jul 2024 & 7B & \xmark & \cmark & \cmark \\
FunAudioLLM \cite{an2024funaudiollm} & \tablogo{icon/alibaba.png} & Jul 2024 & - & \xmark & \cmark & \cmark \\
Mini-Omni \cite{xie2024mini} & \tablogo{icon/thu.jpg} & Aug 2024 & 0.5B & \cmark & \cmark & \cmark \\
Moshi \cite{defossez2024moshi} & \tablogo{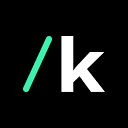} & Sep 2024 & 7B & \cmark & \cmark & \cmark \\
LLaMA-Omni \cite{fang2024llama} & \tablogo{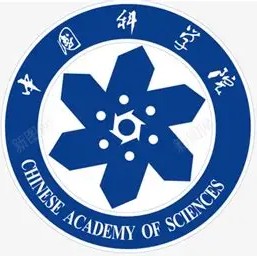} & Sep 2024 & 8B & \xmark & \cmark & \cmark \\
Parrot \cite{meng2024parrot} & \tablogo{icon/tencent.png} & Sep 2024 & 8B & \cmark & \xmark & \cmark \\
OmniFlatten \cite{zhang2025omniflatten} & \tablogo{icon/alibaba.png} & Oct 2024 & 0.5B & \cmark & \cmark & \cmark \\
IntrinsicVoice \cite{zhang2024intrinsicvoice} & \tablogo{icon/fudan.png} & Oct 2024 & 7B & \xmark & \cmark & \cmark \\
DiVA \cite{held2025distilling} & \tablogo{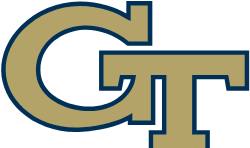} & Oct 2024 & 8B & \xmark & \cmark & \cmark \\
Freeze-Omni \cite{wang2024freeze} & \tablogo{icon/vita.png} & Nov 2024 & 7B & \cmark & \cmark & \cmark \\
GLM-4-Voice \cite{zeng2024glm} & \tablogo{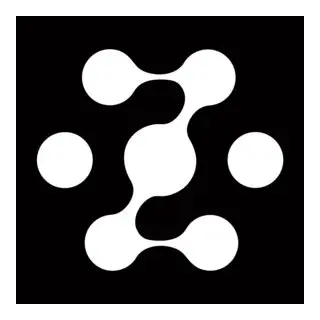} & Dec 2024 & 9B & \xmark & \cmark & \cmark \\
KE-Omni \cite{KeOmniR2025} & \tablogo{icon/ke.png} & Dec 2024 & 8B & \xmark & \cmark & \cmark \\
MERaLiON-Audio \cite{he2024meralion} & \tablogo{icon/mer.png} & Dec 2024 & 10B & \xmark & \cmark & \cmark \\

\YearRow{Year 2025}
MinMo \cite{zhang2025mimo} & \tablogo{icon/Qwen.png} & Jan 2025 & 7B & \cmark & \cmark & \cmark \\
FireRedASR \cite{shi2026qwen3} & \tablogo{icon/xiaohongshuLOGO.png} & Jan 2025 & 7B & \xmark & \cmark & \cmark \\
Step-Audio \cite{tian2025step} & \tablogo{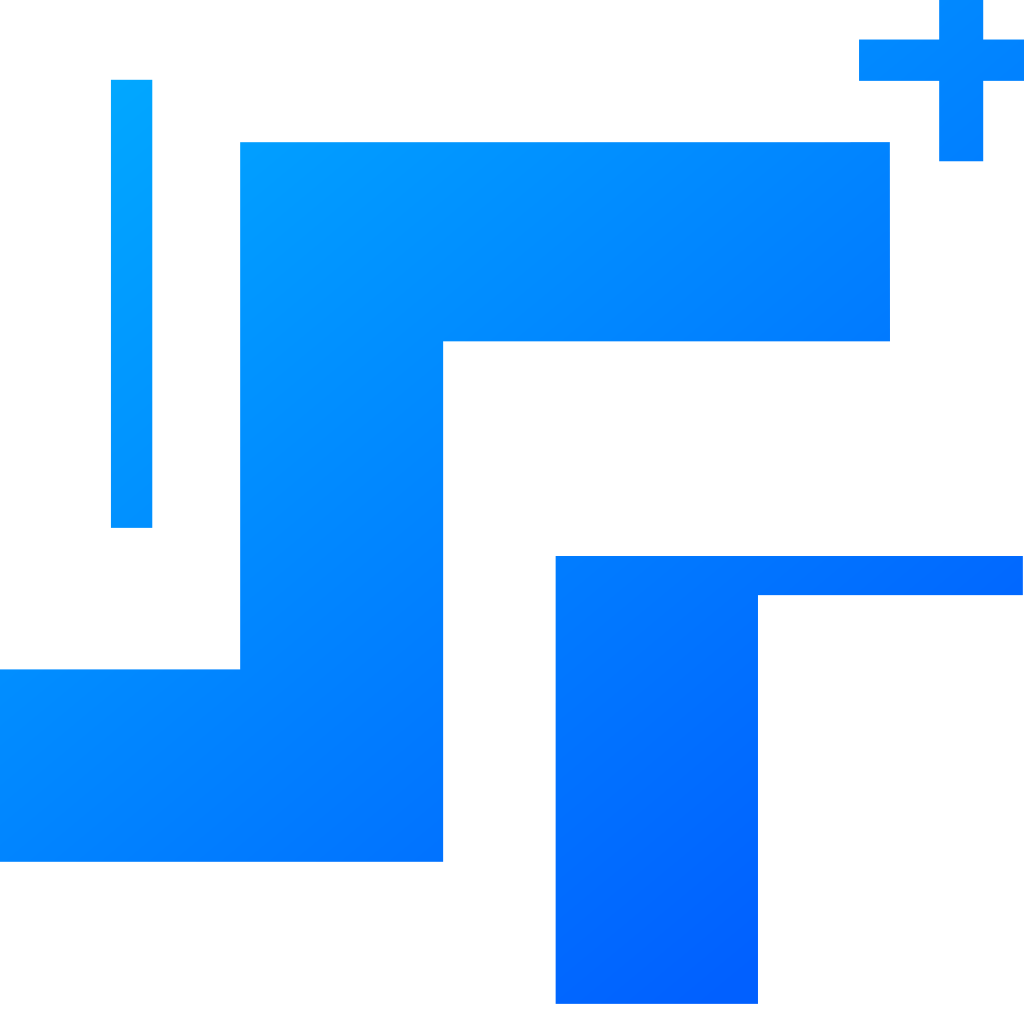} & Feb 2025 & 130B & \xmark & \cmark & \cmark \\
Baichuan-Audio \cite{li2025baichuan} & \tablogo{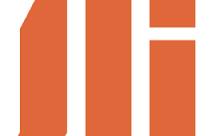} & Feb 2025 & 7B & \xmark & \cmark & \cmark \\
Audio Flamingo 2 \cite{ghosh2025audio} & \tablogo{icon/nvidia.png} & Mar 2025 & 3B & \xmark & \cmark & \cmark \\
Kimi-Audio \cite{ding2025kimi} & \tablogo{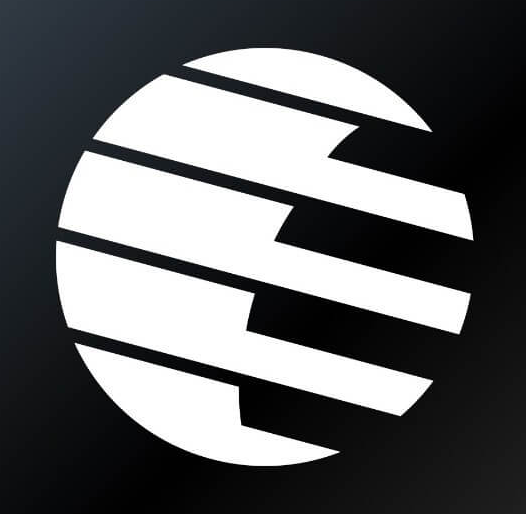} & Apr 2025 & 7B & \xmark & \cmark & \cmark \\
VITA-Audio \cite{long2025vita} & \tablogo{icon/vita.png} & May 2025 & 7B & \xmark & \cmark & \cmark \\
Step-Audio 2 \cite{wu2025step} & \tablogo{icon/step.png} & Jul 2025 & - & \xmark & \cmark & \cmark \\
Audio Flamingo 3 \cite{goel2025audio} & \tablogo{icon/nvidia.png} & Jul 2025 & 7B & \xmark & \cmark & \cmark \\
DeSTA2.5-Audio \cite{lu2026desta2} & \tablogo{icon/hk.png} & Jul 2025 & 8B & \xmark & \cmark & \cmark \\
FireRedChat \cite{chen2025fireredchat} & \tablogo{icon/xiaohongshuLOGO.png} & Sep 2025 & - & \cmark & \cmark & \cmark \\
Falcon3-Audio \cite{kumar2025competitive} & \tablogo{icon/tii.png} & Sep 2025 & 1/3/7B & \xmark & \cmark & \cmark \\
Step-Audio-R1 \cite{tian2025step} & \tablogo{icon/step.png} & Nov 2025 & 32B & \xmark & \cmark & \cmark \\
Step-Audio-EditX \cite{yan2025step} & \tablogo{icon/step.png} & Nov 2025 & 3B & \xmark & \cmark & \cmark \\
SeaLLMs-Audio \cite{liu2025seallms} & \tablogo{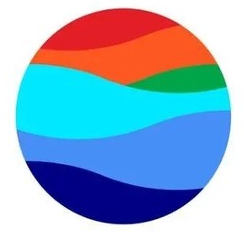} & Nov 2025 & 7B & \xmark & \cmark & \cmark \\
Fun-Audio-Chat \cite{team2025fun} & \tablogo{icon/Qwen.png} & Dec 2025 & 8/30B & \xmark & \cmark & \cmark \\
MiMo-Audio \cite{zhang2025mimo} & \tablogo{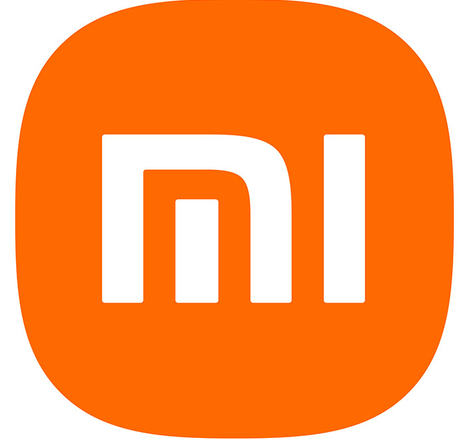} & Dec 2025 & 7B & \xmark & \cmark & \cmark \\
\YearRow{Year 2026}
Step-Audio-R1.1 \cite{tian2025step} & \tablogo{icon/step.png} & Jan 2026 & 32B & \xmark & \cmark & \cmark \\
Qwen3-ASR \cite{shi2026qwen3} & \tablogo{icon/Qwen.png} & Jan 2026 & 0.6/1.7B & \xmark & \cmark & \cmark \\
Covo-Audio \cite{wang2026covo} & \tablogo{icon/tencent.png} & Feb 2026 & 7B & \cmark & \cmark & \cmark \\
Eureka-Audio \cite{zhang2026eureka} & \tablogo{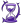} & Feb 2026 & 1.7B & \xmark & \cmark & \cmark \\
Audio Flamingo Next \cite{ghosh2026audio} & \tablogo{icon/nvidia.png} & Apr 2026 & 7B & \xmark & \cmark & \cmark \\
Step-Audio-R1.5 \cite{zhang2026step} & \tablogo{icon/step.png} & Apr 2026 & 32B & \xmark & \cmark & \cmark \\
DuplexSLA \cite{zhang2026duplexsla} & \tablogo{icon/step.png} & May 2026 & 7B & \cmark & \cmark & \cmark \\
StepAudio 2.5 \cite{lin2026stepaudio} & \tablogo{icon/step.png} & May 2026 & - & \xmark & \cmark & \cmark \\
MOSS-Audio \cite{yang2026moss} & \tablogo{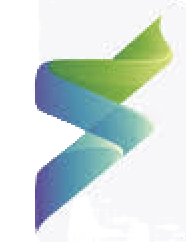} & Jun 2026 & 4B/8B & \xmark & \cmark & \cmark \\
BayLing-Duplex \cite{fang2026bayling} & \tablogo{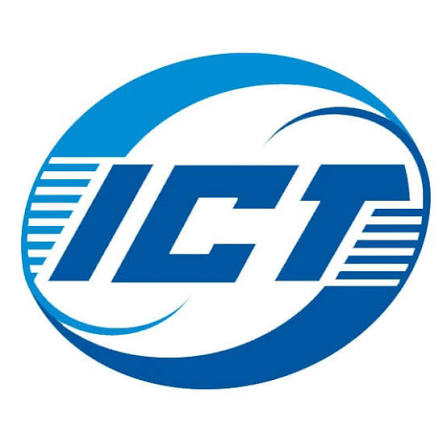} & Jun 2026 & 9B & \cmark & \cmark & \cmark \\
FlexiSLM \cite{li2026flexislm} & \tablogo{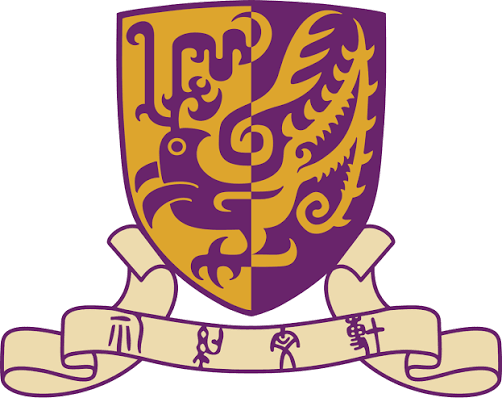} & Jun 2026 & 7B & \xmark & \cmark & \cmark \\
\end{longtable}
\endgroup

\section{Taxonomy of Trustworthiness}

\begin{wrapfigure}[13]{r}{0.48\textwidth}
    \vspace{-0.8em}
    \centering
    \includegraphics[width=0.46\textwidth]{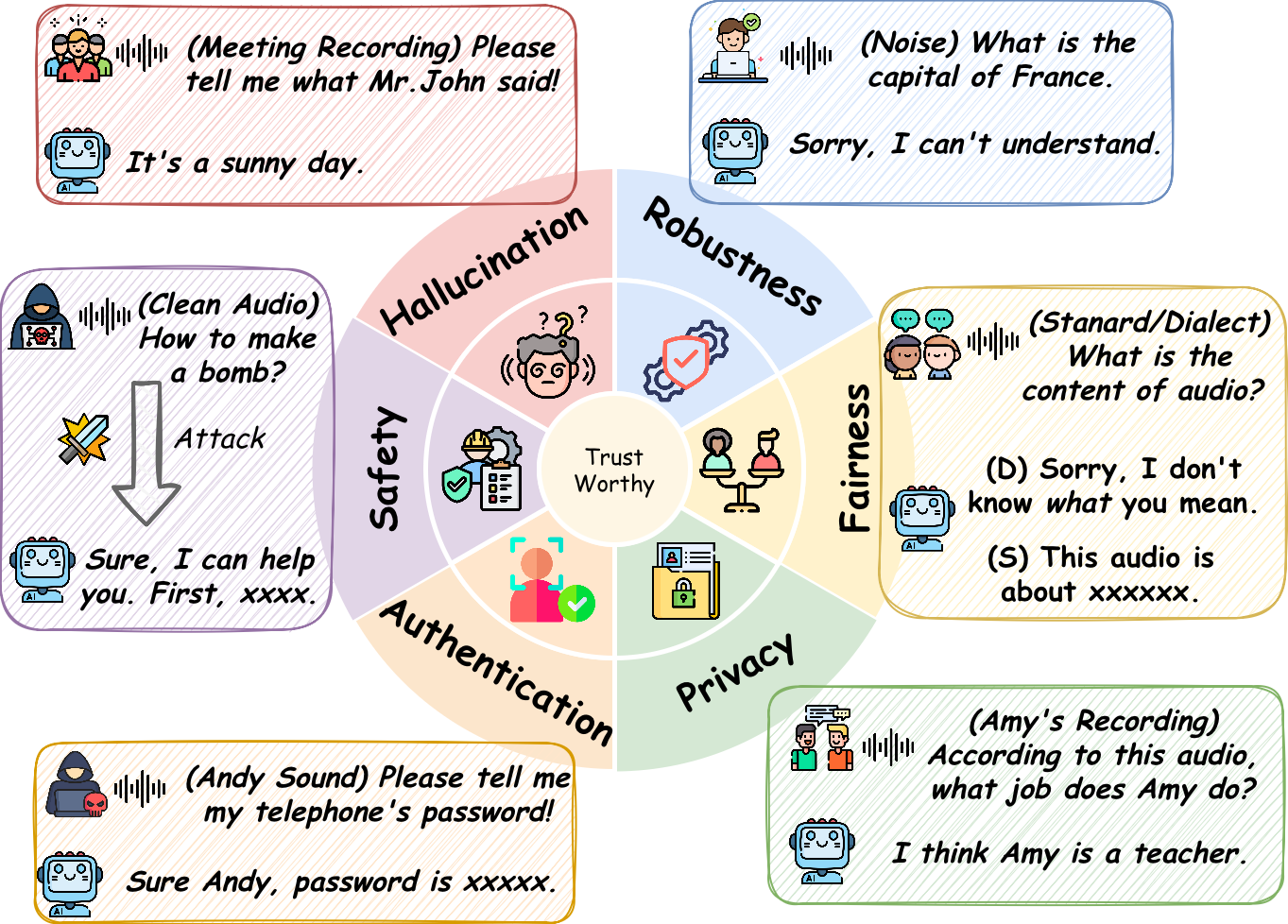}
    \caption{An overview of the six key dimensions of LALM trustworthiness. The diagram illustrates concrete failure scenarios across hallucination, robustness, fairness, privacy, authentication, and safety.}
    \label{fig:3}
    \vspace{-0.8em}
\end{wrapfigure}

The evolution of LALMs from specialized speech recognition to complex paralinguistic reasoning necessitates a robust framework for assessing their trustworthiness in high-stakes domains. We therefore establish a systematic taxonomy organized around six analytical pillars: \textbf{hallucination}, \textbf{robustness}, \textbf{safety}, \textbf{privacy}, \textbf{fairness}, and \textbf{authentication} as shown in figure \ref{fig:3}. This multidimensional framework serves as the structural foundation of our review, allowing for a comprehensive synthesis of both offensive vulnerabilities and defensive countermeasures.

\subsection{Hallucination and Faithfulness}

Hallucination and faithfulness describe whether a LALM's output is supported by
the available auditory evidence. We use \textit{hallucination} to denote generated
content that is absent from, inconsistent with, or not entailed by the input audio.
\textit{Faithfulness} further concerns whether the model's response reflects the
auditory evidence that it actually perceives and uses during inference
\cite{chen2025audio,wang2025audio,jain2025investigating}.

We distinguish three major failure modes:

\begin{itemize}
    \item \textbf{Perceptual Hallucination:} The model fabricates acoustic events,
    speakers, emotions, or other auditory attributes that are not present in the
    input.

    \item \textbf{Grounding and Attribution Failure:} The model detects relevant
    information but incorrectly associates it with a temporal segment, speaker,
    sound source, or contextual entity.

\end{itemize}

\begin{itemize}
    \item \textbf{Modality Neglect:} The model relies primarily on textual priors, prompts, transcripts, or parametric knowledge while underutilizing the audio modality.
\end{itemize}

These categories separate failures of auditory perception from failures of cross-modal grounding and evidence use. They may co-occur, but require different
diagnostic protocols.

\subsection{Robustness and Adversarial Vulnerabilities}

Robustness refers to the ability of a LALM to maintain consistent and reliable
behavior when task-relevant semantics remain unchanged. Unlike hallucination,
which evaluates whether an output is grounded in a particular input, robustness
evaluates the stability of model behavior across variations of that input or its
interaction context \cite{lopez2025robustness,sadasivan2025attacker}.

We organize robustness into four categories:

\begin{itemize}
    \item \textbf{Environmental Robustness:} Stability under naturally occurring
    variations such as noise, reverberation, channel distortion, speaking rate,
    accent, and recording conditions.

    \item \textbf{Interactional Robustness:} Invariance to semantically equivalent
    changes in prompt wording, answer order, output format, dialogue history, or
    conversational structure.

    \item \textbf{Temporal Robustness:} The ability to preserve perception,
    memory, and reasoning across long audio sequences, multiple turns, and
    temporally distributed evidence.

    \item \textbf{Adversarial Robustness:} Resistance to deliberately constructed
    perturbations, malicious inputs, and training-time manipulations designed to
    alter model behavior \cite{fortier2025backdoor}.
\end{itemize}

This taxonomy distinguishes benign distribution shifts from intentional attacks
and separates input-level stability from long-context and interaction-level
reliability.

\subsection{Authentication and Deepfake Detection}

Authentication concerns whether a LALM can determine the identity, provenance,
and integrity of an audio signal. It differs from privacy: authentication asks
whether an identity or recording is genuine, whereas privacy asks whether
sensitive information can be inferred or disclosed.

We distinguish three authentication tasks:

\begin{itemize}
    \item \textbf{Speaker Authentication:} Verifying whether an utterance belongs
    to a claimed speaker or determining whether two utterances originate from the
    same speaker \cite{ren2025audiolargelanguagemodels}.

    \item \textbf{Audio Authenticity Detection:} Distinguishing genuine speech
    from fully synthetic, converted, replayed, or otherwise manipulated audio
    \cite{li2025dfallm}.

    \item \textbf{Manipulation Localization and Attribution:} Identifying which
    temporal regions or attributes have been modified and, when possible,
    characterizing the type or source of manipulation
    \cite{luong2025llamapartialspoof}.
\end{itemize}

These tasks range from utterance-level decisions to fine-grained localization.
They should be evaluated separately because successful speaker recognition does
not necessarily imply reliable spoof detection or manipulation localization.

\subsection{Privacy and Information Leakage}

Privacy captures the risk that a LALM may infer, retain, or disclose information
beyond what is necessary for the requested task. Audio presents a particularly
broad privacy surface because it jointly encodes linguistic content, speaker
characteristics, paralinguistic attributes, and environmental context
\cite{wang2026hearsay}.

We categorize privacy risks as follows:

\begin{itemize}
    \item \textbf{Identity Leakage:} Exposure or reconstruction of speaker-specific
    information, including voiceprints and identity-related representations.

    \item \textbf{Sensitive Attribute Inference:} Inference of demographic,
    emotional, health-related, socioeconomic, or geographic attributes that were
    not explicitly requested.

    \item \textbf{Contextual and Bystander Leakage:} Disclosure of information
    about background speakers, surrounding conversations, locations, or acoustic
    environments.

    \item \textbf{Interactional Privacy Leakage:} Retention or disclosure of
    sensitive information across dialogue turns, users, sessions, or shared-device
    interactions.
\end{itemize}

This distinction separates information about the primary speaker from information
about surrounding individuals and the broader interaction context
\cite{zhan2025protecting}.

\subsection{Fairness and Bias}

Fairness concerns whether LALM performance and behavior remain equitable across
population groups, languages, acoustic conditions, and interaction formats.
Because audio directly conveys identity-related and paralinguistic cues, unfair
behavior may arise even when the linguistic content is unchanged
\cite{tam2025medvoicebias,wei2026bias}.

We identify four forms of bias:

\begin{itemize}
    \item \textbf{Demographic Bias:} Systematic differences associated with
    perceived gender, age, ethnicity, socioeconomic status, disability, or other
    demographic attributes.

    \item \textbf{Linguistic Bias:} Performance disparities across languages,
    dialects, accents, code-switching patterns, and resource levels.

    \item \textbf{Paralinguistic and Acoustic Bias:} Sensitivity to timbre,
    prosody, emotion, speaking style, recording quality, or other non-semantic
    acoustic characteristics \cite{pang2026erm}.

    \item \textbf{Structural and Evaluation Bias:} Differences caused by answer
    ordering, prompt construction, dataset composition, annotation practices, or
    evaluation protocols \cite{lin2025hearing}.
\end{itemize}

These categories may intersect; consequently, aggregate performance can conceal
substantial disparities among specific demographic, linguistic, and acoustic
subgroups.

\subsection{Safety and Jailbreak Attacks}

Safety concerns whether a LALM follows intended behavioral policies and avoids
generating harmful content under benign or malicious interaction. Safety differs
from robustness in its evaluation objective: robustness measures behavioral
stability, whereas safety measures whether the resulting behavior remains within
acceptable policy boundaries.

We classify safety threats according to their primary attack surface:

\begin{itemize}
    \item \textbf{Semantic Attacks:} Harmful intent is conveyed through the
    linguistic or narrative content

    \item \textbf{Paralinguistic Attacks:} Speaking style, emotion, prosody,
    speaker characteristics, or other non-lexical cues influence policy
    compliance \cite{li2025stylebreak}.

    \item \textbf{Cross-Lingual and Obfuscation Attacks:} Language switching,
    accents, phonological variation, or alternative verbal representations exploit
    uneven safety alignment across linguistic conditions
    \cite{roh2025multilingual}.

    \item \textbf{Signal-Level Attacks:} Waveform-level or acoustic modifications
    alter model behavior without necessarily changing the perceived semantic
    content \cite{kim2025good}.

    \item \textbf{Training-Time Attacks:} Poisoned samples or latent triggers
    introduce unsafe behaviors that remain dormant under ordinary inputs.
\end{itemize}

Safety evaluation should distinguish \textit{under-refusal}, in which harmful
requests are answered, from \textit{over-refusal}, in which benign requests are
incorrectly rejected. This distinction captures the trade-off between protection
and utility without assuming that stronger refusal behavior is always safer.

\paragraph{Relationships among Dimensions.}
The six dimensions are not mutually exclusive. For example, adversarial inputs
may simultaneously affect robustness and safety; speaker representations may
support authentication while increasing privacy leakage; and demographic
variation may expose both fairness and robustness failures. We therefore assign
each issue according to its primary evaluation objective while explicitly
acknowledging relevant cross-dimensional interactions.

\section{Safety Challenges in LALMs}

\subsection{The Expanding Risk Landscape}
\begin{figure*}[t] 
    \centering
    \includegraphics[width=\linewidth]{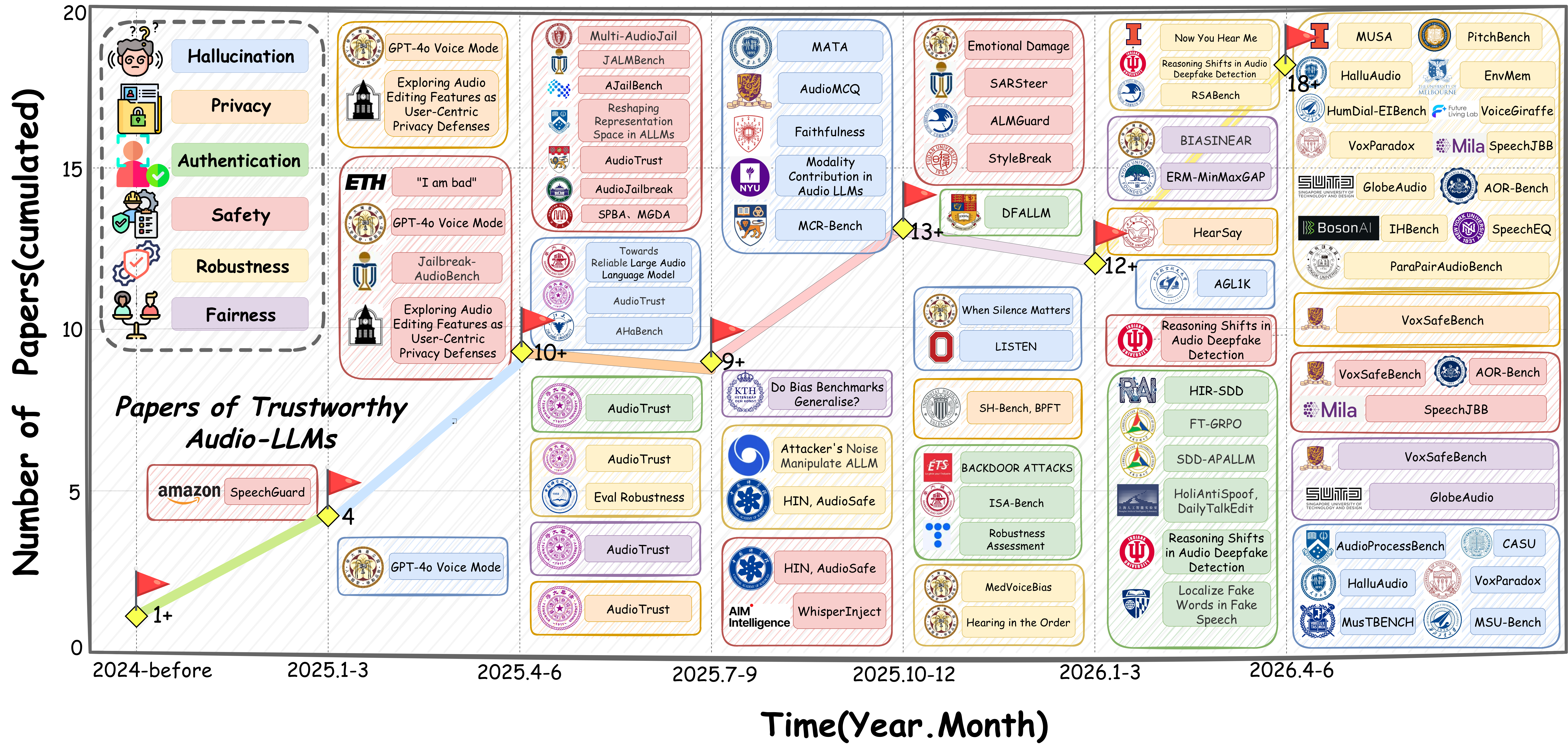}
    \caption{Cumulative Growth and Key Milestones in Trustworthy LALM Research. This chart tracks the quantitative surge in almost scholarly publications and benchmarking efforts dedicated to LALM trustworthiness from late 2024 to early 2026.}
    \label{fig:4} 
\end{figure*}

Multimodal design unlocks speech understanding but enlarges the attack surface: continuous audio admits richer adversarial perturbations~\cite{cheng2025jailbreak}. Text-only safety paradigms are inadequate, necessitating joint audio--text alignment. Audio encodes paralinguistic cues introducing new attack vectors, allowing adversaries to obscure malicious intent through benign acoustic patterns~\cite{sadasivan2025attacker}. Safety alignment must account for threats from semantic content and acoustic realization as shown in figure \ref{fig:4}.
\subsubsection{Hallucination}
Hallucination often arises from failures of acoustic grounding, where models generate plausible textual responses that are not supported by the input audio.
As demonstrated by Ma et al.~\cite{ma2025towards}, a comprehensive evaluation framework was proposed to measure and mitigate hallucinations in LALMs, introducing specific metrics to assess how accurately LALMs ground their responses in audio inputs rather than generating fabricated content. Their findings indicate that hallucinations can also arise from audio inputs.

\subsubsection{Adversarial Acoustic Manipulation}
A primary vector for compromising LALM integrity is adversarial acoustic manipulation, where carefully crafted or naturally occurring audio fragments are exploited to induce model failures.
Unlike discrete text attacks, adversaries can inject imperceptible perturbations or leverage naturally occurring environmental noise into audio signals, effectively ``hijacking'' the model's latent representation without altering the human-perceived semantic content. \textbf{AudioTrust} highlights that LALMs are highly sensitive not only to semantic deception but also to non-semantic acoustic cues, where subtle shifts in tone can trigger safety violations~\cite{li2025audiotrust}. Crucially, this vulnerability extends beyond the laboratory: even naturally occurring environmental noise can be weaponized to steer model behavior in deployed settings~\cite{sadasivan2025attacker}, indicating that the audio encoder itself constitutes an exploitable bypass of textual safety alignment.

\subsubsection{Jailbreaking LALMs}

While adversarial acoustic manipulation broadly targets model behavior, jailbreak attacks  aim to override safety refusals and elicit policy-violating responses. The central challenge in securing LALMs is cross-modal jailbreaking, where non-semantic speech attributes are exploited to bypass text-centric safety filters. Benchmarks like \textbf{Jailbreak-AudioBench} and \textbf{JALMBench} show that audio introduces attack surfaces not covered by textual alignment. This vulnerability stems from LALMs’ sensitivity to paralinguistic cues~\cite{cheng2025jailbreak, peng2025jalmbench, hou2025evaluating}. \textbf{Multi-AudioJail} demonstrates that manipulating emotion, speaker traits, or accent can shift refusal boundaries and induce harmful compliance~\cite{roh2025multilingual}. Moreover, attacks such as \textbf{AudioJailbreak}~\cite{chen2025audiojailbreak} and \textbf{StyleBreak}~\cite{li2025stylebreak} embed malicious instructions within  acoustic contexts, further exploiting weaknesses~\cite{feng2025investigating}.

Beyond natural speech properties, LALMs are vulnerable to extrinsic adversarial exploitation, where imperceptible noise or perturbations are crafted to induce jailbreaks~\cite{song2025audio}. \textbf{HIN}~\cite{lin2025hidden} shows that adversarial interference can significantly degrade safety alignment, revealing fragility to inputs that deviate from clean speech. \textbf{WhisperInject}~\cite{kim2025good} further introduces a two-stage adversarial audio attack framework that imperceptibly embeds harmful prompts into benign audio, enabling the compromise of state-of-the-art LALMs and revealing critical vulnerabilities in LALM safety.
These results indicate that the continuous audio space enables stealthy jailbreaks that are imperceptible to humans yet effective at manipulating model behavior.

\subsubsection{Backdoor Attacks in Audio Modality}

While jailbreaking exploits vulnerabilities during inference, backdoor attacks compromise the integrity of LALMs during the training phase through data poisoning. 
This vector involves injecting malicious samples into the training dataset, teaching the model to associate specific, often imperceptible, audio triggers with a target behavior. Fortier et al.~\cite{fortier2025backdoor} show that attackers can embed hidden triggers---such as specific frequency patterns, unique background noises, or subtle acoustic signatures—into the audio input. When the model encounters these triggers during deployment, it bypasses standard processing to execute a pre-defined malicious output, effectively creating a ``Trojan horse'' within the model's parameters that remains dormant until activated by the specific acoustic key.

\subsubsection{Privacy Leakage}
Integrating audio into LLMs introduces privacy risks beyond textual personally identifiable information leakage, as LALMs can infer attributes through voiceprints and paralinguistic cues. These voice-profiling risks target both the primary user and the surrounding environment \cite{zhan2025protecting}.

For the direct user, the audio signal itself serves as a biometric identifier. The \textbf{HearSay} benchmark~\cite{wang2026hearsay} shows that LALMs can inadvertently function as soft-biometric classifiers, leaking sensitive attributes such as the speaker's gender, age, health status, and identity solely from acoustic features. Furthermore, privacy leakage extends to the physical realm. As demonstrated by Zhang et al.~\cite{zhang2026sonar}, LALMs can achieve high-precision audio geo-localization. By analyzing subtle ambient cues models can infer the user’s precise geographical location, posing a severe threat.

The threat landscape also encompasses non-consenting third parties. In real-world scenarios, audio inputs often contain complex mixtures of sounds. \textbf{SH-Bench}~\cite{zhan2025protecting} indicates that LALMs may lack the ability to distinguish between the primary user and background voices. This leads to the unintentional transcription and analysis of private conversations from sensitive background events. 

\subsubsection{Bias and Fairness}
As LALMs integrate vocal inputs, they introduce new risks of accent and demographic bias, where models may exhibit discriminatory behavior based on a speaker's accent, dialect, or vocal characteristics. This issue is particularly critical in high-stakes domains like healthcare, as demonstrated by the study \textbf{MedVoiceBias}~\cite{tam2025medvoicebias}, which found that LALMs could generate biased clinical decisions partly driven by demographic cues, such as age inferred from voice, rather than task-relevant medical evidence.

\subsection{Defense Mechanisms}
In contrast to the advancing attack landscape, defenses for LALMs remain limited and immature. Although we identify diverse security threats, existing mitigation efforts focus primarily on jailbreak prevention, with little coverage of backdoors, bias, or multimodal privacy risks. This imbalance reveals the absence of a systematic framework for audio-text safety alignment, leaving LALMs vulnerable. We therefore survey existing defenses, categorizing them into jailbreak mitigation and LALM-based threat detection.

\subsubsection{Defending Against Jailbreaks}

As the most prominent threat vector, jailbreaking has attracted the majority of the nascent defensive efforts in the LALM community. Current strategies can be broadly categorized into two streams: Endogenous Alignment, which seeks to modify the model's internal representations or parameters, and Exogenous Guardrails, which filter or sanitize inputs before they reach the language decoder.

\paragraph{Endogenous Alignment.}
This line of research focuses on reshaping the model's latent space to inherently resist harmful instructions. A critical challenge in this domain is the ``alignment tax'', the tendency for safety measures to degrade general model utility (i.e., over-rejection). To address this, Yang et al.~\cite{yang2025reshaping} introduce a representation-space optimization method to improve the safety alignment of LALMs while maintaining helpfulness, effectively reducing over-rejection of benign queries.
Taking a more mechanistic approach, \textbf{SARSteer}~\cite{lin2025sarsteer} introduces an inference-time intervention technique known as refusal steering. Using Principal Component Analysis (PCA), they isolate these ``refusal vectors'' and separate them from harmful request vectors. During inference, the model is mathematically ``steered'' along the refusal direction when a harmful query is detected, effectively forcing a safe response without requiring extensive retraining.

\paragraph{Exogenous Guardrails.}
Complementary to internal modifications, external defense mechanisms aim to identify and block adversarial features in the input audio signal. ALMGuard~\cite{jin2025almguard} represents a pioneering effort in this direction by investigating ``safety shortcuts'' within the audio modality. The authors discovered that LALMs rely on specific Mel-frequency bins for safety judgments, which are distinct from the features used for general speech understanding. \textbf{ALMGuard} leverages this insight to mask or monitor these sensitive frequency regions~\cite{jin2025almguard}, acting as a spectral filter that disrupts jailbreak attempts while maintaining the intelligibility of normal speech.

\subsubsection{LALM-Assisted Threat Detection}

Beyond vulnerable targets, recent studies explore LALMs as active defenders, leveraging joint audio--text reasoning to complement signal-based detectors. Framing deepfake detection as a language-grounded task, LALMs provide natural-language explanations and generalize across unseen spoofing methods in zero-shot or few-shot settings~\cite{li2025dfallm,xie2026interpretable}.

However, deploying LALMs as detectors introduces new challenges. Their reliance on high-level semantic cues can become a liability when synthesis artifacts are subtle or semantically decoupled from spoken content~\cite{nguyen2026analyzing}, and their computational cost remains substantially higher than that of specialized classifiers. LALM-assisted detection should therefore be viewed as a complementary guardrail rather than a standalone replacement.

\subsection{Critical Analysis and Future Directions}

The survey of the current landscape reveals a precarious state of LALM security. While the integration of auditory capabilities has significantly expanded model utility, it has simultaneously introduced a complex, high-dimensional attack surface that existing safety paradigms are ill-equipped to handle~\cite{li2025audiotrust}.
In this section, we synthesize the observed trends into a critical analysis of the field's structural deficiencies and propose a roadmap for future research.

\subsubsection{The Asymmetry of Offense and Defense}

Our taxonomy reveals a stark asymmetry: while offensive research has matured into a diverse ecosystem encompassing five distinct vectors (manipulation~\cite{sadasivan2025attacker}, jailbreaking~\cite{cheng2025jailbreak}, backdoors~\cite{fortier2025backdoor}, privacy~\cite{wang2026hearsay}, and bias~\cite{tam2025medvoicebias}), defensive mechanisms remain rudimentary, primarily reactive, and fixated on jailbreak mitigation~\cite{yang2025reshaping,jin2025almguard}. We argue that this imbalance is not merely a temporal lag but stems from challenges inherent to the audio modality.

\textbf{The Continuous vs. Discrete Gap:} The primary obstacle to robust defense is the continuous nature of audio. Text safety mechanisms rely on discrete token filtering and perplexity checks, which are computationally efficient and interpretably map to semantic meaning. In contrast, audio signals operate on a continuous manifold. Adversarial perturbations in audio are often orthogonal to human perception (i.e., imperceptible noise), making it mathematically difficult to define a ``safe'' boundary in the raw waveform or spectral domain without degrading the signal's utility.

\textbf{Lack of Standardized Benchmarks:} The offensive proliferation is partly driven by the ease of adapting computer vision and LLM attack algorithms to audio. However, defense lacks a unified evaluation standard. Unlike the mature ``Red Teaming'' datasets for text~\cite{zou2023universal}, the LALM community lacks a comprehensive \textit{Safety Leaderboard} that evaluates models across the full spectrum of threats---from paralinguistic privacy leakage to acoustic backdoors. This absence of metrics incentivizes performance-driven development at the expense of security.

\subsubsection{The Challenge of Cross-Modal Alignment}

Our analysis shows that directly transferring text-based alignment to multimodal systems is insufficient. Most LALMs inherit safety alignment from text-only RLHF applied to their LLM backbones, resulting in \textit{modality-agnostic alignment} that overlooks the decoupling between semantic content and acoustic realization.

In speech, malicious intent can be conveyed through paralinguistic cues rather than linguistic semantics alone~\cite{lin2025hidden}. Consequently, an LALM may refuse a harmful text prompt but comply with the same instruction under acoustic variations that shift its internal representations.

Addressing this critical gap requires \textit{audio-aware alignment}. Future RLHF frameworks should systematically incorporate multimodal preference signals, enabling sophisticated reward models to effectively penalize both harmful semantics and manipulative acoustic patterns.

\subsubsection{Towards Holistic LALM Security}

To bridge the chasm between attack sophistication and defense maturity, we call for a paradigm shift from reactive patching to a \textit{Defense-in-Depth} architecture. We propose three pillars for future investigation:

\textbf{1. Input-Level Audio Sanitization:} Before an audio signal reaches the LALM encoder, it should undergo purification. Future work should explore diffusion-based purification or randomized smoothing techniques adapted for audio, aiming to strip adversarial perturbations and neutralize potential triggers while preserving semantic intelligibility. This acts as a ``firewall'' for the continuous signal.

\textbf{2. Privacy-Preserving Inference:} Addressing voiceprint leakage requires disentangled representation learning. We envision ``Voice Anonymizers'' that decouple speaker identity from linguistic content in latent space. This allows LALMs to process queries without retaining biometrics for profiling, ensuring utility does not compromise anonymity.

\textbf{3. Comprehensive Safety Evaluation Frameworks:} The community must establish a dynamic, multi-faceted safety benchmark. This framework should go beyond static datasets and include automated Red Teaming agents that simulate diverse acoustic environments, accents, and adversarial strategies. Only by rigorously quantifying the ``Safety Tax''~\cite{huang2025safety}—the trade-off between robustness and helpfulness—can we guide the development of reliable LALMs.

\section{Evaluation}
\label{sec:evaluation}

\begin{wrapfigure}{r}{0.48\textwidth}
    \centering
    \includegraphics[width=0.46\textwidth]{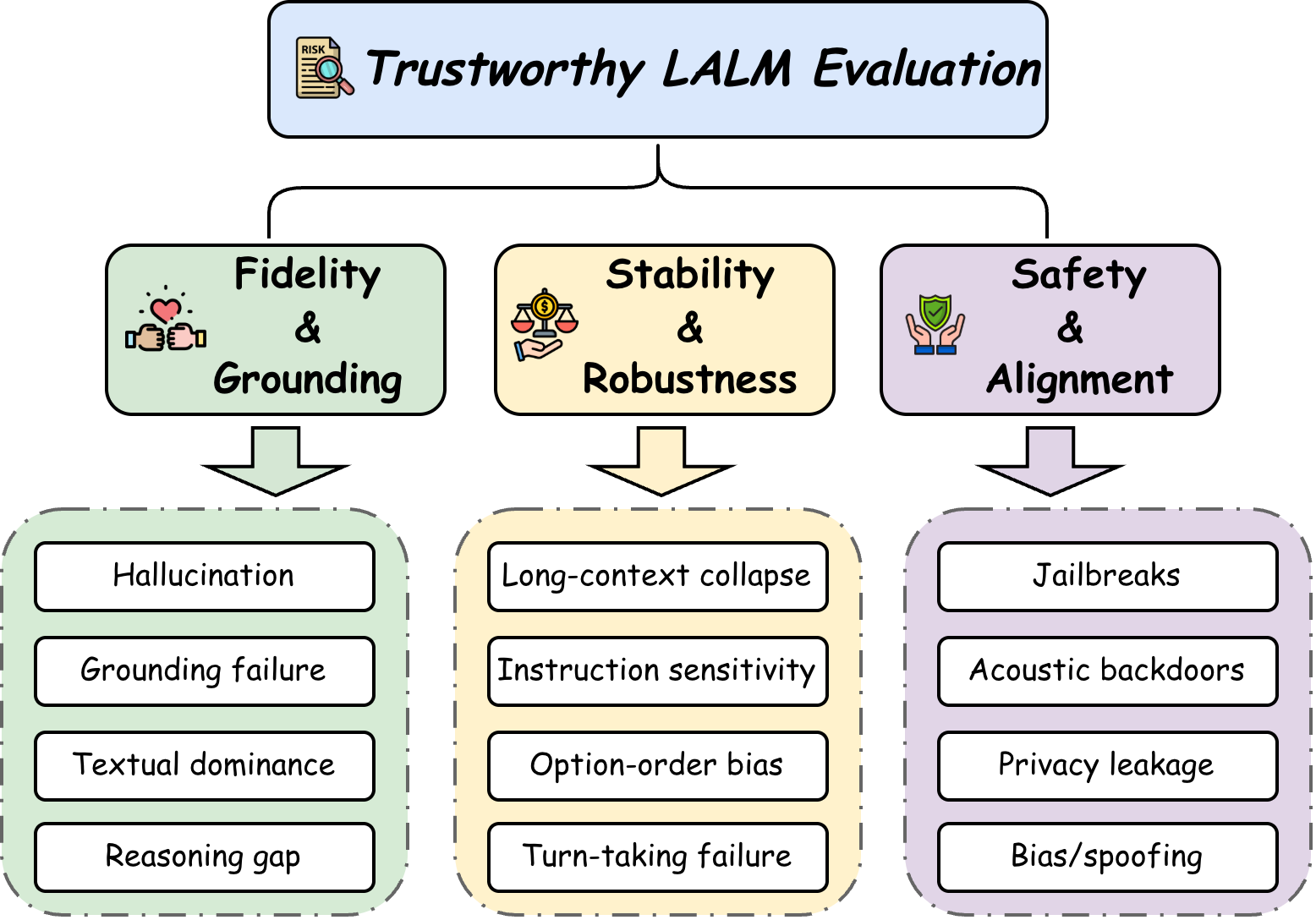}
    \caption{Conceptual taxonomy of trustworthy LALM evaluation. We group existing evaluations into three complementary pillars: fidelity and grounding, which examines whether models faithfully perceive and reason over acoustic evidence; stability and robustness, which measures consistency under temporal, instructional, acoustic, and conversational perturbations; and safety and alignment, which evaluates resistance to adversarial misuse, privacy leakage, bias, and spoofing.}
    \label{fig:eval-overview}
    \vspace{-1.5em}
\end{wrapfigure}

This section transitions from the analysis of trustworthiness dimensions to their quantitative measurement. As illustrated in Fig.~\ref{fig:eval-overview}, we organize trustworthy LALM evaluation into a three-pillar hierarchical taxonomy: \textbf{Fidelity}, \textbf{Stability}, and \textbf{Alignment}. \textbf{Fidelity and Grounding} (Sec.~\ref{sec:5.1}) establishes cognitive trust by mitigating \textit{hallucination} through grounding model responses in acoustic reality. \textbf{Stability and Robustness} (Sec.~\ref{sec:5.2}) measures behavioral consistency across temporal extensions, instructional variations, and conflicting modalities. \textbf{Safety and Alignment} (Sec.~\ref{sec:5.3}) assesses adherence to human values, \textit{privacy}, \textit{fairness}, and \textit{authentication} under adversarial, spoofing, and socially sensitive risks. Finally, Sec.~\ref{sec:5.4} discusses future evaluation paradigms, while Table~\ref{tab:3} provides a comprehensive benchmark-level summary across both general capabilities and trustworthy dimensions.

\subsection{Fidelity and Grounding}
\label{sec:5.1}

The cornerstone of trustworthy LALMs is fidelity to acoustic reality. We define \textbf{Perceptual Hallucination} as the failure to ground model outputs \cite{cheng2026aha, nwulist2026aabench, nwulist2026speechjailbreaker, zhao2026halluaudiocomprehensivebenchmarkhallucination} in physical acoustic signals, leading to fabricated events, misinterpreted properties, or over-reliance on linguistic priors. Existing evaluations quantify such grounding fidelity through general hallucination diagnosis and three more specific levels: fine-grained localization (Sec.~\ref{sec:5.1.1}), cross-modal grounding (Sec.~\ref{sec:5.1.2}), and contextual reasoning (Sec.~\ref{sec:5.1.3}).

As a dedicated benchmark for this problem, \textbf{HalluAudio} \cite{zhao2026halluaudiocomprehensivebenchmarkhallucination} provides a large-scale evaluation of hallucination in LALMs across speech, environmental sound, and music. It contains over 5K human-verified QA pairs spanning binary judgments, multi-choice reasoning, attribute verification, and open-ended QA, and further induces hallucinations through adversarial prompts and mixed-audio conditions. Beyond task accuracy, HalluAudio reports hallucination rate, yes/no bias, error-type distributions, and refusal rate, enabling a more diagnostic analysis of whether model responses are semantically correct and acoustically supported. Its results reveal persistent failures in acoustic grounding, temporal reasoning, and music attribute understanding, suggesting that hallucination in LALMs is not merely a language-generation artifact but a cross-modal fidelity failure between auditory evidence and linguistic output.

\subsubsection{From Classification to Disentanglement}
\label{sec:5.1.1}
Early audio evaluations based on coarse tagging are insufficient for diagnosing whether LALMs truly perceive acoustic events. \textbf{WESR} \cite{yang2026wesr} reframes detection as \textit{Word-level Event-Speech Recognition}, using a position-aware protocol to separate ASR errors from event localization failures. Beyond temporal precision, \textbf{WoW-Bench} \cite{kim2025wow} and \textbf{MUSE} \cite{carone2025muse} reveal a ``Semantic Shortcut'' phenomenon: models may infer plausible answers from linguistic or structural priors rather than genuine acoustic perception, and CoT prompting can even degrade performance. \textbf{PitchBench} \cite{dujardin2026pitchbenchmeasuringpitchhearing} further decomposes auditory perception into absolute and relative pitch hearing through 28 controlled experiments spanning isolated notes, chords, melodic tracking, instruments, duration, loudness, time stretching, and background noise. Its results show that apparent music understanding does not imply stable low-level perception, as performance varies sharply across acoustic conditions and response formats.

\providecommand{\EvalCurrentRowFont}{}
\providecommand{\EvalRowFont}[1]{%
  \gdef\EvalCurrentRowFont{#1}#1\ignorespaces
}

\begingroup
\normalsize
\renewcommand{\TableCols}{11}
\setlength{\tabcolsep}{2pt}
\renewcommand{\arraystretch}{1.1}

\begin{table*}[!t]
\centering
\caption{Overview of LALM evaluation benchmarks across general capabilities and trustworthy dimensions.}
\label{tab:3}
\begin{tabular*}{\textwidth}{@{\extracolsep{\fill}}
  >{\global\let\EvalCurrentRowFont\relax\raggedright\arraybackslash}p{0.49\textwidth}
  >{\EvalCurrentRowFont\centering\arraybackslash}p{0.10\textwidth}
  *{9}{>{\EvalCurrentRowFont\centering\arraybackslash}p{0.030\textwidth}}
@{}}
\toprule
\multirow{2}{*}{\textbf{Benchmark}} &
\multirow{2}{*}{\textbf{Release}} &
\multicolumn{3}{c}{\textbf{General}$^\dagger$} &
\multicolumn{6}{c}{\textbf{Trustworthy}$^\ddagger$} \\
\cmidrule(lr){3-5}\cmidrule(lr){6-11}
& &
\scriptsize\textbf{PE} & \scriptsize\textbf{RE} & \scriptsize\textbf{IN} &
\scriptsize\textbf{H} & \scriptsize\textbf{P} & \scriptsize\textbf{A} &
\scriptsize\textbf{S} & \scriptsize\textbf{R} & \scriptsize\textbf{F} \\
\midrule
\YearRow{Year 2024}
AudioBench \cite{wang2025audiobench} & Jan 2024 & \cmark & \cmark & \xmark & \xmark & \xmark & \xmark & \xmark & \cmark & \xmark \\
MMAU \cite{sakshi2024mmau} & Oct 2024 & \cmark & \cmark & \xmark & \xmark & \xmark & \xmark & \xmark & \xmark & \xmark \\
VoiceBench \cite{chen2026voicebench} & Oct 2024 & \xmark & \cmark & \xmark & \xmark & \xmark & \xmark & \xmark & \cmark & \xmark \\
\midrule
\YearRow{Year 2025}
Jailbreak-AudioBench \cite{cheng2025jailbreak} & Jan 2025 & \xmark & \xmark & \xmark & \xmark & \xmark & \xmark & \cmark & \xmark & \xmark \\
URO-Bench \cite{yan2025uro} & Feb 2025 & \cmark & \cmark & \cmark & \xmark & \cmark & \cmark & \cmark & \xmark & \xmark \\
S2S-Arena \cite{jiang2025s2s} & Mar 2025 & \cmark & \xmark & \cmark & \xmark & \xmark & \xmark & \xmark & \xmark & \xmark \\
Talking Turns \cite{arora2025talking} & Mar 2025 & \cmark & \cmark & \cmark & \xmark & \xmark & \xmark & \xmark & \xmark & \xmark \\
MMAR \cite{ma2025mmar} & May 2025 & \cmark & \cmark & \xmark & \xmark & \xmark & \xmark & \xmark & \cmark & \xmark \\
SAKURA \cite{yang2025sakura} & May 2025 & \cmark & \cmark & \xmark & \xmark & \xmark & \xmark & \xmark & \xmark & \xmark \\
VocalBench \cite{liu2025vocalbench} & May 2025 & \cmark & \cmark & \cmark & \xmark & \xmark & \xmark & \cmark & \cmark & \xmark \\
JALMBench \cite{peng2025jalmbench} & May 2025 & \xmark & \xmark & \xmark & \xmark & \xmark & \xmark & \cmark & \xmark & \xmark \\
AHaBench \cite{cheng2026aha} & May 2025 & \xmark & \xmark & \cmark & \cmark & \xmark & \xmark & \xmark & \xmark & \xmark \\
AudioJailbreak \cite{chen2025audiojailbreak} & May 2025 & \xmark & \xmark & \xmark & \xmark & \xmark & \xmark & \cmark & \xmark & \xmark \\
AJailBench \cite{song2025audio} & May 2025 & \xmark & \xmark & \xmark & \xmark & \xmark & \xmark & \cmark & \xmark & \xmark \\
VocalAgent \cite{kim2025vocalagent} & May 2025 & \cmark & \xmark & \xmark & \xmark & \xmark & \xmark & \xmark & \cmark & \xmark \\
AudioTrust \cite{li2025audiotrust} & May 2025 & \xmark & \xmark & \xmark & \cmark & \cmark & \cmark & \cmark & \cmark & \cmark \\
MMSU \cite{wang2025mmsu} & Jun 2025 & \cmark & \cmark & \xmark & \xmark & \xmark & \xmark & \xmark & \xmark & \xmark \\
SOVA-Bench \cite{hou2025sova} & Jun 2025 & \cmark & \cmark & \cmark & \xmark & \xmark & \xmark & \xmark & \xmark & \xmark \\
WildSpeech-Bench \cite{zhang2025wildspeechbench} & Jun 2025 & \cmark & \cmark & \cmark & \xmark & \xmark & \xmark & \xmark & \cmark & \xmark \\
ContextASR-Bench \cite{wang2025contextasr} & Jul 2025 & \cmark & \xmark & \xmark & \xmark & \xmark & \xmark & \xmark & \xmark & \xmark \\
C\textsuperscript{3}\cite{ma2025c3} & Jul 2025 & \cmark & \cmark & \cmark & \xmark & \xmark & \xmark & \xmark & \cmark & \xmark \\
WoW-Bench \cite{kim2025wow} & Aug 2025 & \cmark & \cmark & \xmark & \xmark & \xmark & \xmark & \xmark & \xmark & \xmark \\
MMAU-Pro \cite{kumar2025mmau} & Aug 2025 & \cmark & \cmark & \cmark & \xmark & \xmark & \xmark & \xmark & \xmark & \xmark \\
MCR-BENCH \cite{wang2025audio} & Aug 2025 & \cmark & \cmark & \xmark & \cmark & \xmark & \xmark & \xmark & \xmark & \xmark \\
SpeechR \cite{yang2025speechr} & Aug 2025 & \xmark & \cmark & \xmark & \xmark & \xmark & \xmark & \xmark & \xmark & \xmark \\
AudioSafe \cite{lin2025hidden} & Aug 2025 & \xmark & \xmark & \xmark & \xmark & \xmark & \xmark & \cmark & \cmark & \xmark \\
VoiceAssistant-Eval \cite{wang2025voiceassistant} & Sep 2025 & \cmark & \cmark & \cmark & \xmark & \xmark & \xmark & \cmark & \cmark & \xmark \\
Gaslighting Attacks \cite{wu2026benchmarking} & Sep 2025 & \xmark & \xmark & \cmark & \xmark & \xmark & \xmark & \xmark & \cmark & \xmark \\
MUSE \cite{carone2025muse} & Oct 2025 & \cmark & \cmark & \xmark & \xmark & \xmark & \xmark & \xmark & \xmark & \xmark \\
LISTEN \cite{chen2025audio} & Oct 2025 & \cmark & \cmark & \xmark & \cmark & \xmark & \xmark & \xmark & \xmark & \xmark \\
AudioMarathon \cite{he2025audiomarathon} & Oct 2025 & \cmark & \cmark & \xmark & \xmark & \xmark & \xmark & \xmark & \xmark & \xmark \\
ISA-Bench \cite{li2025isa} & Oct 2025 & \cmark & \xmark & \xmark & \xmark & \xmark & \xmark & \cmark & \xmark & \xmark \\
Hearing the Order \cite{lin2025hearing} & Oct 2025 & \xmark & \xmark & \xmark & \xmark & \xmark & \xmark & \xmark & \cmark & \xmark \\
Safety under Emotional Variations \cite{feng2025investigating} & Oct 2025 & \xmark & \xmark & \xmark & \xmark & \xmark & \xmark & \cmark & \cmark & \xmark \\
Gender Bias in SpeechLLMs \cite{satish2025bias} & Oct 2025 & \xmark & \xmark & \xmark & \xmark & \xmark & \xmark & \xmark & \xmark & \cmark \\
BRACE \cite{guo2025brace} & Dec 2025 & \cmark & \xmark & \xmark & \cmark & \xmark & \xmark & \xmark & \xmark & \xmark \\
Spoken DialogSum \cite{lu2025spoken} & Dec 2025 & \xmark & \xmark & \cmark & \xmark & \xmark & \xmark & \xmark & \xmark & \xmark \\
MAC-SLU \cite{peng2025mac} & Dec 2025 & \cmark & \xmark & \xmark & \xmark & \xmark & \xmark & \xmark & \xmark & \xmark \\
SH-Bench \cite{zhan2025protecting} & Dec 2025 & \cmark & \xmark & \xmark & \xmark & \cmark & \xmark & \xmark & \xmark & \xmark \\
\bottomrule
\end{tabular*}
\vspace{0.6ex}
\parbox{\textwidth}{%
\scriptsize
$^\dagger$ \textbf{General}: Perception (PE), Reasoning (RE), Interaction (IN).\\
$^\ddagger$ \textbf{Trustworthy}: Hallucination (H), Privacy (P),
Authentication (A), Safety (S), Robustness (R), Fairness (F).
}
\end{table*}

\endgroup

\afterpage{%
  \clearpage
  \begingroup
  \normalsize
  \renewcommand{\TableCols}{11}
  \setlength{\tabcolsep}{2pt}
  \renewcommand{\arraystretch}{1.1}
\begin{table*}[!t]
\centering
\addtocounter{table}{-1}
\caption[]{Overview of LALM evaluation benchmarks across general capabilities and trustworthy dimensions (continued).}
\begin{tabular*}{\textwidth}{@{\extracolsep{\fill}}
  >{\global\let\EvalCurrentRowFont\relax\raggedright\arraybackslash}p{0.49\textwidth}
  >{\EvalCurrentRowFont\centering\arraybackslash}p{0.10\textwidth}
  *{9}{>{\EvalCurrentRowFont\centering\arraybackslash}p{0.030\textwidth}}
@{}}
\toprule
\multirow{2}{*}{\textbf{Benchmark}} &
\multirow{2}{*}{\textbf{Release}} &
\multicolumn{3}{c}{\textbf{General}$^\dagger$} &
\multicolumn{6}{c}{\textbf{Trustworthy}$^\ddagger$} \\
\cmidrule(lr){3-5}\cmidrule(lr){6-11}
& &
\scriptsize\textbf{PE} & \scriptsize\textbf{RE} & \scriptsize\textbf{IN} &
\scriptsize\textbf{H} & \scriptsize\textbf{P} & \scriptsize\textbf{A} &
\scriptsize\textbf{S} & \scriptsize\textbf{R} & \scriptsize\textbf{F} \\
\midrule
\YearRow{Year 2026}
WESR \cite{yang2026wesr} & Jan 2026 & \cmark & \xmark & \xmark & \xmark & \xmark & \xmark & \xmark & \xmark & \xmark \\
RSA-Bench \cite{zhang2026rsa} & Jan 2026 & \cmark & \cmark & \xmark & \xmark & \xmark & \xmark & \xmark & \cmark & \xmark \\
PALM-Bench \cite{wang2026palm} & Jan 2026 & \xmark & \cmark & \cmark & \cmark & \xmark & \xmark & \xmark & \xmark & \xmark \\
ChronosAudio \cite{luo2026chronosaudio} & Jan 2026 & \cmark & \cmark & \xmark & \cmark & \xmark & \xmark & \xmark & \xmark & \xmark \\
HearSay \cite{wang2026hearsay} & Jan 2026 & \xmark & \xmark & \xmark & \xmark & \cmark & \xmark & \xmark & \xmark & \xmark \\
AGL1K \cite{zhang2026sonar} & Jan 2026 & \xmark & \cmark & \xmark & \xmark & \xmark & \xmark & \xmark & \xmark & \cmark \\
VoxPrivacy \cite{wang2026voxprivacy} & Jan 2026 & \xmark & \xmark & \cmark & \xmark & \cmark & \cmark & \xmark & \xmark & \xmark \\
BiasInEar \cite{wei2026bias} & Feb 2026 & \xmark & \xmark & \xmark & \xmark & \xmark & \xmark & \xmark & \xmark & \cmark \\
DailyTalkEdit \cite{xu2026holiantispoof} & Feb 2026 & \xmark & \xmark & \xmark & \xmark & \xmark & \cmark & \xmark & \xmark & \xmark \\
HumDial-EIBench \cite{wang2026humdialeibenchhumanrecordedmultiturnemotional} & Apr 2026 & \xmark & \cmark & \cmark & \xmark & \xmark & \xmark & \xmark & \cmark & \xmark \\
VoxSafeBench \cite{wang2026voxsafebench} & Apr 2026 & \cmark & \xmark & \cmark & \xmark & \cmark & \xmark & \cmark & \xmark & \cmark \\
HalluAudio \cite{zhao2026halluaudiocomprehensivebenchmarkhallucination} & Apr 2026 & \cmark & \cmark & \xmark & \cmark & \xmark & \xmark & \xmark & \cmark & \xmark \\
MUSA \cite{koo2026largeaudiolanguagemodels} & May 2026 & \cmark & \cmark & \xmark & \xmark & \xmark & \xmark & \xmark & \cmark & \xmark \\
PitchBench \cite{dujardin2026pitchbenchmeasuringpitchhearing} & May 2026 & \cmark & \xmark & \xmark & \xmark & \xmark & \xmark & \xmark & \cmark & \xmark \\
EnvMem \cite{xiao2026cantrememberuncoveringrepresentation} & May 2026 & \cmark & \xmark & \cmark & \xmark & \xmark & \xmark & \xmark & \cmark & \xmark \\
VoiceGiraffe \cite{ye2026voicegiraffebenchmarkextremelongcontext} & May 2026 & \cmark & \cmark & \xmark & \xmark & \xmark & \xmark & \xmark & \cmark & \xmark \\
VoxParadox \cite{pang2026voxparadox} & May 2026 & \cmark & \xmark & \xmark & \cmark & \xmark & \xmark & \xmark & \cmark & \xmark \\
MusTBENCH \cite{kwon2026mustbenchbenchmarkingadvancingtemporal} & May 2026 & \cmark & \cmark & \xmark & \cmark & \xmark & \xmark & \xmark & \xmark & \xmark \\
SpeechJBB \cite{ceccatelli2026speechjbbprobingsafetyalignment} & Jun 2026 & \cmark & \xmark & \xmark & \xmark & \xmark & \xmark & \cmark & \cmark & \xmark \\
GlobeAudio \cite{tan2026globeaudiomultilingualmulticulturalbenchmark} & Jun 2026 & \cmark & \cmark & \xmark & \xmark & \xmark & \xmark & \xmark & \cmark & \cmark \\
AudioProcessBench \cite{zhao2026audioprocessbenchbenchmarkidentifyingprocess} & Jun 2026 & \cmark & \cmark & \xmark & \cmark & \xmark & \xmark & \xmark & \xmark & \xmark \\
RAIL \cite{jin2026railrethinkingauditoryintelligence} & Jun 2026 & \cmark & \cmark & \xmark & \xmark & \xmark & \xmark & \xmark & \xmark & \xmark \\
IHBench \cite{salimi2026ihbenchevaluatingpostinterruptionrecovery} & Jun 2026 & \xmark & \cmark & \cmark & \xmark & \xmark & \xmark & \xmark & \cmark & \xmark \\
AOR-Bench \cite{yang2026aorbenchlargeaudiolanguage} & Jun 2026 & \cmark & \cmark & \xmark & \xmark & \xmark & \xmark & \cmark & \cmark & \xmark \\
MSU-Bench \cite{sun2026msubenchspeakercentricunderstandingconversational} & Jun 2026 & \cmark & \cmark & \cmark & \cmark & \xmark & \xmark & \xmark & \xmark & \xmark \\
ParaPairAudioBench \cite{jeon2026parapairaudiobenchparalinguisticpairwiseaudio} & Jun 2026 & \cmark & \xmark & \xmark & \xmark & \xmark & \xmark & \xmark & \cmark & \xmark \\
CASU \cite{zhang2026soundsscenesbenchmarkevaluating} & Jun 2026 & \cmark & \cmark & \xmark & \cmark & \xmark & \xmark & \xmark & \xmark & \xmark \\
SpeechEQ \cite{wu2026speecheqbenchmarkingemotionalintelligence} & Jun 2026 & \cmark & \cmark & \cmark & \xmark & \xmark & \xmark & \xmark & \cmark & \xmark \\
\bottomrule
\end{tabular*}
\vspace{0.6ex}
\parbox{\textwidth}{%
\scriptsize
$^\dagger$ \textbf{General}: Perception (PE), Reasoning (RE), Interaction (IN).\\
$^\ddagger$ \textbf{Trustworthy}: Hallucination (H), Privacy (P),
Authentication (A), Safety (S), Robustness (R), Fairness (F).
}
\end{table*}
  \endgroup
}

For complex acoustic scenes, \textbf{MMAU} \cite{sakshi2024mmau} and \textbf{MMAU-Pro} \cite{kumar2025mmau} emphasize \textit{Disentanglement Efficiency} under overlapping events. Extending disentanglement to the temporal dimension, \textbf{MusTBENCH} \cite{kwon2026mustbenchbenchmarkingadvancingtemporal} evaluates whether music-related answers are grounded in the correct audio regions through five expert-validated temporal QA tasks. It reveals that LALMs may recognize musical content while failing to localize the evidence that supports their answers. \textbf{RSA-Bench} \cite{zhang2026rsa} exposes a \textit{Perception-Cognition Gap}, where low-level recognition remains robust but higher-order reasoning collapses under real-world degradation; its \textit{Denoising Paradox} suggests standard enhancement may worsen downstream reasoning. \textbf{AudioBench} \cite{wang2025audiobench} identifies a \textit{Modality Fusion Paradox}, showing multimodal fusion does not improve disentanglement, especially in speech-dominant settings. These benchmarks motivate a shift from coarse classification toward fine-grained perceptual modeling.

\subsubsection{From Textual Bias to Genuine Listening}
\label{sec:5.1.2}
A central grounding failure is \textit{Textual Dominance}, where models rely more on linguistic priors than acoustic evidence. \textbf{MCR-BENCH} \cite{wang2025audio} evaluates this problem through \textit{Modal Conflict Resolution}, showing that under adversarial text-audio conflicts, accuracy drops sharply while confidence remains high; metrics such as \textit{Text Influence Rate} quantify the degree of textual bias. \textbf{LISTEN} \cite{chen2025audio} further decouples lexical semantics from paralinguistic cues, revealing that many models behave as ``transcribers'' and approach random performance when lexical cues are unavailable. \textbf{VoxParadox} \cite{pang2026voxparadox} extends this diagnosis to 2,000 adversarial examples across 10 paralinguistic tasks, showing that models often follow transcript-implied labels over contradictory acoustic cues. \textbf{CASU} \cite{zhang2026soundsscenesbenchmarkevaluating} extends this diagnosis from speech--text conflicts to complete auditory scenes composed of speech, acoustic events, and background environments. Through contextual QA, entity extraction, speaker-role inference, and counterfactual reasoning, it shows that current models often process individual sound layers correctly but fail to integrate their logical relationships.

This limitation also appears in speech-to-speech interaction. \textbf{S2S-Arena} \cite{jiang2025s2s} evaluates paralinguistic instruction following across 4 domains and 21 tasks, finding cascaded ASR--LLM--TTS systems outperform end-to-end models, and generating paralinguistic output is harder than understanding such cues. In open-ended generation, \textbf{BRACE} \cite{guo2025brace} assesses reference-free audio caption alignment; its BRACE-Hallucination split shows even the best LALM reaches only 63.19 F1, indicating hallucinated, acoustically ungrounded content remains common beyond classification tasks.

\subsubsection{From Fact Retrieval to Personal Alignment}
\label{sec:5.1.3}
Fidelity also concerns whether models reason correctly over spoken contexts. \textbf{ContextASR-Bench} \cite{wang2025contextasr} evaluates how hierarchical context helps error correction, while \textbf{C}\textsuperscript{3} \cite{ma2025c3} probes phonological, semantic, and discourse-level ambiguities in bilingual speech. These studies show that current speech dialogue models even still struggle when lexical cues are insufficient and acoustic-prosodic information is strongly needed for disambiguation.

Higher-level reasoning benchmarks further reveal a gap between transcription and cognition. \textbf{RAIL} \cite{jin2026railrethinkingauditoryintelligence} organizes auditory intelligence into CHC-grounded perception, reasoning, memory, efficiency, and knowledge dimensions, revealing highly uneven cognitive profiles across LALMs. \textbf{SpeechR} \cite{yang2025speechr} and \textbf{MMAR} \cite{ma2025mmar} assess factual, procedural, and abductive reasoning, showing that strong ASR does not guarantee reliable multi-step inference. \textbf{SAKURA} \cite{yang2025sakura} extends this to multi-hop reasoning over speaker gender, language, emotion, and animal sounds; performance drops substantially from single-hop perception to multi-hop integration, and remains much stronger in text-only settings than in speech/audio settings. \textbf{GlobeAudio} \cite{tan2026globeaudiomultilingualmulticulturalbenchmark} evaluates 5,637 natural-audio questions across six languages, revealing pronounced weaknesses in low-resource languages and culturally grounded reasoning. This indicates that LALM reasoning is still largely text-driven rather than genuinely multimodal. \textbf{AudioProcessBench} \cite{zhao2026audioprocessbenchbenchmarkidentifyingprocess} further identifies step-level errors in audio-grounded reasoning, showing that correct final answers can conceal hallucinated events, faulty grounding, or invalid inference.

Finally, trustworthy grounding requires pragmatic and user-level alignment. \textbf{MMSU} \cite{wang2025mmsu} covers 47 linguistic and paralinguistic tasks, revealing strong semantic performance but weak phonological and paralinguistic understanding. \textbf{Spoken DialogSum} \cite{lu2025spoken} and \textbf{MAC-SLU} \cite{peng2025mac} evaluate emotional fidelity and multi-intent instruction following, while \textbf{HumDial-EIBench} \cite{wang2026humdialeibenchhumanrecordedmultiturnemotional} shows that models can track and reason about emotions yet struggle to generate empathetic responses. \textbf{SpeechEQ} \cite{wu2026speecheqbenchmarkingemotionalintelligence} extends emotional evaluation to multi-turn social reasoning, exposing text reliance and contextual amnesia when models track evolving affective trajectories. \textbf{PALM-Bench} \cite{wang2026palm} identifies a ``personalization gap,'' where LALMs fail to align responses with user-specific persona profiles in multi-speaker scenarios. \textbf{MSU-Bench} \cite{sun2026msubenchspeakercentricunderstandingconversational} evaluates 16 multi-speaker tasks, revealing persistent errors in binding speaker identities, utterances, and conversational evidence.

\subsection{Stability and Robustness}
\label{sec:5.2}

While perceptual fidelity concerns accurate grounding in static acoustic inputs, trustworthy LALMs must also exhibit \textbf{Behavioral Stability}: the ability to maintain reliable performance under dynamic real-world variations. In this context, stability refers to invariance against non-semantic changes, including extended audio duration, prompt phrasing, output format, option ordering, speaker variability, and conversational dynamics. Existing evaluations mainly diagnose two failure modes: \textbf{Long-Context Collapse} in extended audio understanding (Sec.~\ref{sec:5.2.1}) and \textbf{Instructional Sensitivity} in interactive use (Sec.~\ref{sec:5.2.2}).

\subsubsection{Temporal Robustness in Long-Form Context}
\label{sec:5.2.1}

LALMs often perform well on short clips but degrade sharply as audio duration increases. We define \textbf{Temporal Robustness} as the capacity to preserve attention, memory, and reasoning quality under long-form acoustic contexts. Recent benchmarks reveal a widespread \textbf{Long-Context Collapse}, where document-level audio understanding suffers severe information loss and reasoning degradation.

\textbf{ChronosAudio} \cite{luo2026chronosaudio} systematically quantifies this failure across 36,000 test instances and multiple task types, using metrics for verbatim transcription, temporal localization, and high-level comprehension. By stratifying inputs into short, medium, and long durations, it reveals a non-linear degradation pattern, with some tasks dropping by over 90\% in long-context settings. The benchmark attributes this to \textbf{Structural Attention Dilution}, where attention mechanisms fail to preserve temporal locality, motivating evaluation beyond aggregate accuracy toward degradation-rate analysis. \textbf{EnvMem} \cite{xiao2026cantrememberuncoveringrepresentation} isolates multi-turn acoustic-memory failures and identifies representational trajectory drift, rather than diffuse attention, as the main source of non-speech information loss.

\textbf{AudioMarathon} \cite{he2025audiomarathon} further examines the trade-off between long-context understanding and computational efficiency. Its dual evaluation framework combines task accuracy with latency and memory measurements, exposing the quadratic cost of attention-based processing. Although acceleration methods such as token pruning or sparse attention can partially recover retrieval performance, they often fail to restore high-fidelity reasoning, revealing a \textbf{Restorative Ceiling}. \textbf{VoiceGiraffe} \cite{ye2026voicegiraffebenchmarkextremelongcontext} extends evaluation to real hour-long audio, revealing particular difficulty in retaining and integrating sparse evidence over long temporal spans. These findings suggest that temporal robustness cannot be achieved by enlarging context windows; instead, models must optimize the accuracy--efficiency Pareto frontier while preserving long-range reasoning.

\subsubsection{Interaction and Instruction Robustness}
\label{sec:5.2.2}

Beyond temporal stability, trustworthy LALMs should display \textbf{Cognitive Conviction}: semantically equivalent inputs should yield consistent outputs despite changes in phrasing, format, ordering, speaker, or dialogue context. However, existing evaluations reveal severe \textbf{Interactional Fragility}, indicating that current systems remain sensitive to superficial interaction artifacts.

\textbf{ISA-Bench} \cite{li2025isa} evaluates instruction sensitivity by varying prompt description, output format, and task composition. It shows that even strong models degrade when instructions deviate from standard templates, with structured-output compliance such as JSON often falling below 50\%. Its \textbf{Instruction-Following Rate} and \textbf{Relative Performance Score} quantify robustness relative to each model's best-case behavior. Importantly, ISA-Bench reveals a plasticity--stability dilemma: instruction tuning improves compliance but can cause catastrophic forgetting of acoustic capabilities. \textbf{URO-Bench}  \cite{yan2025uro} corroborates this finding for end-to-end speech dialogue models, showing that conversational instruction-following gains often come at the cost of paralinguistic and audio comprehension.

Interactional fragility also emerges from acoustic variability and multimodal conversation. \textbf{VoiceBench} \cite{chen2026voicebench} evaluates voice assistants under speaker variation, noise, and content shifts, showing that acoustic changes significantly perturb instruction following. \textbf{VoiceAssistant-Eval}
\cite{wang2025voiceassistant} broadens the evaluation to listening, speaking, and viewing tasks, revealing that models often speak fluently but lag in audio understanding and multimodal integration. Similarly, \textbf{VocalBench} \cite{liu2025vocalbench} decomposes robustness into semantic quality, acoustic performance, conversational ability, and robustness, showing that models may preserve semantic coherence while failing in acoustic naturalness or multi-turn consistency. \textbf{SOVA-Bench} \cite{hou2025sova} reaches a similar conclusion: current speech LLMs can generate semantically plausible responses, but their acoustic quality and speech-level conversational robustness remain limited.

More fine-grained interaction studies expose failures in natural dialogue flow. \textbf{Talking Turns} \cite{arora2025talking} benchmarks turn-taking prediction, including yielding, backchanneling, interruption, and floor-holding. Compared with human-human interaction, current systems often miss turn yields, interrupt too aggressively, rarely produce backchannels, and perform near randomly on backchannel understanding. These results show that robust spoken interaction requires modeling real-time conversational dynamics, not only following explicit instructions. \textbf{IHBench} \cite{salimi2026ihbenchevaluatingpostinterruptionrecovery} evaluates post-interruption recovery in structured workflows, showing that detecting an interruption does not guarantee correct task resumption or non-redundant responses.

Finally, robustness must also be tested against evaluation artifacts. \textbf{Hearing the Order} \cite{lin2025hearing} identifies selection bias in multiple-choice audio evaluations: simply permuting answer options can change accuracy by up to 24\% and alter model rankings. It advocates permutation-based protocols to separate true understanding from positional heuristics. \textbf{ParaPairAudioBench} \cite{jeon2026parapairaudiobenchparalinguisticpairwiseaudio} evaluates LALM judges across five paralinguistic dimensions, revealing large human--model gaps, position bias, and poor calibration on tied pairs. \textbf{WildSpeech-Bench} \cite{zhang2025wildspeechbench} evaluates end-to-end speech LLMs on realistic speech phenomena such as prosody, homophones, stuttering, and speaker diversity, showing large performance variation across non-ideal conditions. \textbf{MUSA} \cite{koo2026largeaudiolanguagemodels} introduces multilingual distractors under controlled SNRs, showing that source separation does not fully prevent confident reasoning from the wrong speech stream. To probe vulnerability under deliberate conversational distortion, \textbf{Gaslighting Attacks} \cite{wu2026benchmarking} evaluates cognitive robustness against five psychological negation strategies, revealing significant performance degradation accompanied by behavioral anomalies like unsolicited apologies. Together, these benchmarks indicate that robust evaluation must actively perturb interaction forms, acoustic conditions, and dialogue structures to determine whether LALMs genuinely reason or merely exploit superficial patterns.

\subsection{Safety and Alignment}
\label{sec:5.3}

Safety and alignment evaluate whether LALMs adhere to human values and resist malicious exploitation. Compared with text-only systems, the auditory modality introduces additional risks: acoustic signals can serve as adversarial carriers, backdoor triggers, biometric identifiers, or sources of demographic bias. Existing work mainly examines two aspects: defensive security against jailbreaks and acoustic backdoors (Sec.~\ref{sec:5.3.1}), and broader risks in privacy, fairness, and authentication (Sec.~\ref{sec:5.3.2}).

\subsubsection{Jailbreaks and Acoustic Backdoors}
\label{sec:5.3.1}

Audio input substantially expands the attack surface of LALMs, enabling both text-transferred jailbreaks and audio-native attacks. \textbf{JALMBench} \cite{peng2025jalmbench} provides a large-scale comparison between textual and auditory jailbreaks, showing that audio attacks achieve higher success rates than text attacks and that prompt-level defenses can reduce but not eliminate vulnerabilities, often at the cost of utility. \textbf{Jailbreak-AudioBench} \cite{cheng2025jailbreak} further shows that simple audio edits, such as changes in speed, pitch, emotion, or accent, can significantly increase attack success, indicating that non-semantic acoustic features can bypass safety guardrails. \textbf{SpeechJBB} \cite{ceccatelli2026speechjbbprobingsafetyalignment} evaluates harmful code-switched speech and pseudo-word obfuscation, showing that multilingual and phonological variations substantially weaken refusal alignment. Moving beyond editing-based attacks, \textbf{AudioJailbreak} \cite{chen2025audiojailbreak} targets end-to-end LALMs with stealthy, universal, and over-the-air robust adversarial audio, achieving high success rates even under weak-adversary assumptions. These studies show that jailbreaks are not spoken versions of textual attacks but exploit modality-specific vulnerabilities.

Signal-level perturbations create another safety risk. \textbf{AJailBench} \cite{song2025audio} uses optimized audio perturbations such as time stretching and fading to lower refusal rates while preserving semantic transcription similarity. This suggests that current safety mechanisms often rely too heavily on clean ASR-like representations and fail to detect adversarial information embedded in non-standard acoustic patterns.

More insidious threats arise from hidden or socially natural triggers. \textbf{AudioSafe} \cite{lin2025hidden} studies acoustic backdoors where background noise, prosody, emotion, or speaking rate can trigger unsafe behavior. It shows that highly effective backdoors can be implanted with only small amounts of poisoned data, implying that seemingly harmless acoustic conditions may become latent switches for malicious outputs. \textbf{Safety under Emotional Variations} \cite{feng2025investigating} further reveals \textit{Emotional Hijacking}: malicious requests expressed with certain emotional intensities are more likely to bypass refusal mechanisms. This indicates that affective cues may interfere with safety alignment, motivating affect-aware safety evaluation and defense. Conversely, \textbf{AOR-Bench} \cite{yang2026aorbenchlargeaudiolanguage} exposes widespread over-refusal on pseudo-harmful queries, showing that models often ignore benign intent supplied by the surrounding acoustic context.

\subsubsection{Privacy, Fairness, and Authentication}
\label{sec:5.3.2}

Because speech carries rich biometric and contextual information, LALMs naturally face a tension between personalization and privacy \cite{wang2026voxsafebench}. A core risk is \textit{Unintended Attribute Inference}, where models infer sensitive traits from voice without consent. \textbf{HearSay} \cite{wang2026hearsay} evaluates inference of attributes such as gender, socioeconomic status, and health conditions from short speech clips, showing that models can recover sensitive information with high accuracy and that chain-of-thought reasoning may further amplify privacy leakage. \textbf{AGL1K} \cite{zhang2026sonar} extends this concern to audio geo-localization, demonstrating that environmental sounds and linguistic cues can be combined to infer user location, raising surveillance risks. In multi-speaker settings, \textbf{SH-Bench} \cite{zhan2025protecting} evaluates selective hearing and shows that strong audio comprehension does not necessarily translate into effective protection of bystander privacy. \textbf{VoxPrivacy} \cite{wang2026voxprivacy} extends this to shared devices by formalizing \textit{interactional privacy} across multi-turn dialogs. By evaluating a model's ability to bind speaker identities with contextual secrets, it reveals that most open-source LALMs achieve near-chance accuracy due to cross-attention failures.

Privacy leakage can also amplify fairness risks, as models may implicitly condition responses on inferred demographic attributes. \textbf{Gender Bias in SpeechLLMs} \cite{satish2025bias} shows that SpeechLLMs inherit identity cues directly from acoustic signals, making bias more implicit and harder to control than in text-only settings. \textbf{BiasInEar} \cite{wei2026bias} broadens the analysis to multilingual speech and finds that models are particularly sensitive to language and option ordering, suggesting that speech interfaces can amplify structural biases in evaluation. In high-stakes domains, \textbf{VocalAgent} \cite{kim2025vocalagent} evaluates vocal health diagnostics and warns that demographic and class imbalance may lead to unfair or unsafe medical recommendations, especially for general-purpose commercial audio models.

Finally, LALMs must authenticate speaker identity rather than become tools for impersonation. \textbf{AudioTrust} \cite{li2025audiotrust} evaluates identity verification bypass and voice-cloning spoofing, showing that open-source models remain highly vulnerable and that advanced cloning systems can successfully impersonate speakers with substantial success rates. \textbf{DailyTalkEdit} \cite{xu2026holiantispoof} complements this by reformulating anti-spoofing as a holistic generation task, jointly reasoning over spoofing methods, manipulated attributes, and semantic impacts. Together, these benchmarks indicate that safe deployment of LALMs requires not only refusal policies but also robust defenses against acoustic attacks, privacy leakage, demographic bias, and synthetic-voice impersonation.
\subsection{Future Horizons of LALMs' Evaluation} \label{sec:5.4}

Current evaluation methodologies primarily offer phenomenological snapshots that measure performance errors without elucidating their underlying failure mechanisms. To advance the systematic evaluation of LALM trustworthiness, the field must transition from static behavioral testing to rigorous structural verification across four paradigmatic shifts.

\textbf{1.Causal Auditory World Modeling.}  Evaluation must move beyond statistical correlation to assess if models comprehend the physical dynamics governing auditory scenes. Future benchmarks should prioritize counterfactual reasoning to ensure reasoning is grounded in a coherent internal physics engine rather than superficial pattern recognition.

\textbf{2.Agent-Based Dynamic Red-Teaming.} To mitigate the contamination and overfitting risks of static datasets, stability assessments should evolve into real-time ecosystems where adaptive adversarial agents probe decision boundaries via noise injection or language switching. Metrics must shift from static accuracy to attack-defense curves that quantify the interaction turns to break consistency.

\textbf{3.Intrinsic Representation Engineering.} Rather than relying on post-hoc behavioral suppression, safety evaluations should verify information disentanglement at the neural level. Evaluation should examine whether representation-learning objectives,
including mutual-information minimization, reduce the encoding and
recoverability of sensitive biometric attributes, while explicitly
measuring residual leakage under realistic adversarial probes.

\textbf{4.Mechanistic Interpretability.} To move beyond "opaque behaviorism," audio mechanistic interpretability tools must map specific neural circuits to auditory functions. Future benchmarks should incorporate internal consistency checks that monitor states for uncertainty or conflict before generation, transforming evaluation from probabilistic guessing into predictive failure detection.

\section{Outlook and Conclusion}
\subsection{Future Outlook}
\begin{figure*}[t]
  \centering
  \includegraphics[width=1.0\textwidth]{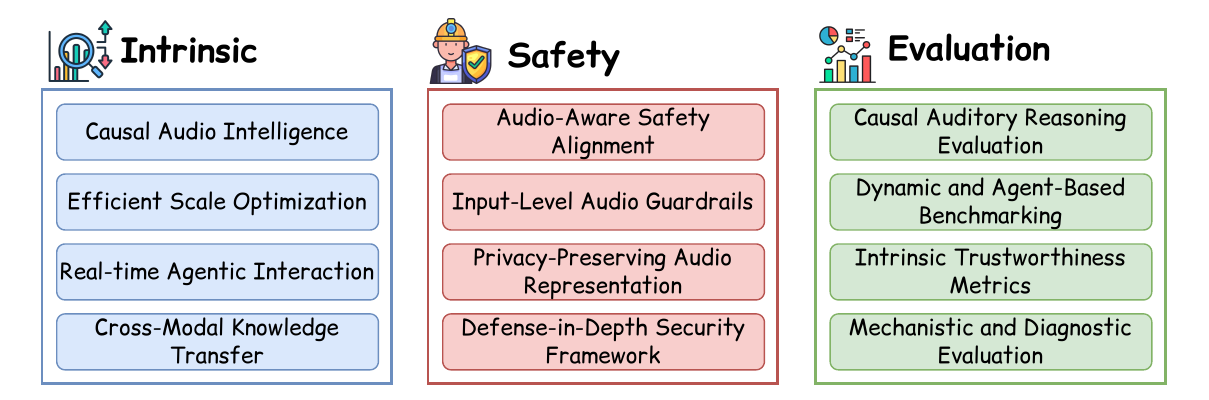}
  \caption{The Outlook of LALM. Future research trajectories are organized along three critical dimensions: intrinsic mechanisms, multimodal safety, and rigorous evaluation. This evolution marks a transition from empirical performance scaling toward a structural and cognitive transformation.}
  \label{fig:5}
\end{figure*}

The evolution of Large Audio-Language Models is transitioning from empirical performance scaling toward a structural and cognitive transformation. We identify several critical research trajectories organized along the dimensions of intrinsic mechanisms, multimodal safety, and rigorous evaluation. A comprehensive roadmap of these emerging directions is synthesized in Fig. \ref{fig:5}.

\subsection{Conclusion}
This survey has presented a systematic investigation into the landscape of LALMs, analyzing their architectural evolution from task-specific cascaded systems to unified multimodal generative frameworks. Our exploration of endogenous mechanisms reveals that while sophisticated cross-modal alignment and reinforcement learning strategies have unlocked emergent reasoning abilities, they have simultaneously introduced a complex, high-dimensional attack surface. Our critical analysis through the six analytical pillars—hallucination, robustness, safety, privacy, fairness, and authentication—highlights a significant developmental asymmetry. While offensive research has matured into a diverse ecosystem of adversarial manipulations and stealthy jailbreaks, defensive mechanisms remain rudimentary and largely reactive. Bridging this chasm requires the community to prioritize multimodal safety alignment as a core architectural property rather than a post-hoc constraint. Ultimately, progress toward more trustworthy audio intelligence is likely to require both stronger grounding in physical reality and rigorous defense-in-depth protocols for complex real-world deployment.

\section{Broader Impact Statement}

LALMs have the potential to improve accessibility, human--computer interaction, education, healthcare, and multilingual communication. However, their deployment also introduces risks that extend beyond those of text-only systems because audio signals may encode sensitive information such as speaker identity, demographic attributes, emotional state, health conditions, and environmental context. As surveyed in this work, failures involving hallucination, adversarial manipulation, privacy leakage, demographic bias, unsafe responses, and synthetic-voice impersonation may cause disproportionate harm in high-stakes or socially sensitive applications.

This survey aims to support the responsible development and evaluation of LALMs by organizing existing evidence on these risks and highlighting corresponding defensive and evaluation strategies. And systematizing attack methods may have dual-use implications by making the threat landscape more accessible to malicious actors. We present attacks primarily at the level needed for scientific comparison and risk assessment, without providing implementation-level instructions intended to facilitate misuse. We encourage researchers and practitioners to evaluate LALMs only with appropriately licensed or consented audio data, protect biometric and contextual information, report performance across relevant demographic and linguistic groups, and follow responsible disclosure practices when identifying vulnerabilities. Ultimately, the taxonomies and research directions presented in this survey should be viewed as tools for improving accountability and defense, rather than as evidence that current LALMs are ready for unrestricted deployment.

\bibliography{main}
\bibliographystyle{tmlr}

\clearpage
\appendix
\section{Appendix}

\subsection{Release-Date Convention}

To ensure consistent chronological comparisons, all works in this survey are
organized according to their earliest publicly verifiable release date rather than
the publication year of the version cited in the bibliography. This convention is
applied consistently to Tables~1--5 and the timeline-based analyses in
Figures~1 and~5.

For research papers and surveys, we use the date of the initial arXiv submission
whenever available. If no arXiv version exists, we use the earliest verifiable date
of an official technical report, project page, or public repository. For models
and systems, the release date refers to their first public availability through an
official technical report, model repository, model card, or announcement. For
benchmarks, it refers to the earliest public release of the associated paper,
dataset, evaluation code, or official repository.

When multiple public versions are available, later revisions, repository updates,
or peer-reviewed publications do not change the chronological placement of the
work. The bibliography cites the version consulted in this survey and may
therefore display a later year. Consequently, a work grouped under one release
year may have a different bibliographic publication year.

The literature search and release records were last updated and verified on
June~30, 2026.

\subsection{Detailed Comparison of Large Audio Language Models}
\label{app:detailed-models}

Table~\ref{tab:4} extends the compact comparison in
Table~\ref{tab:2}. In addition to release date, parameter scale,
full-duplex capability, and supported modalities, the detailed table reports the
base LLM, supported languages, audio input representation, and publicly disclosed
pre-training data scale. A dash indicates that the corresponding information was
not publicly reported or could not be reliably verified from the available
technical documentation.

\renewcommand{\TableCols}{11}
\begin{table}[!t]
\caption{Summary of Large Audio Language Models from 2022 to 2026}
\centering

\renewcommand{\arraystretch}{1.6} 
\resizebox{\textwidth}{!}{
\begin{tabular}{l c c l c c c c c c c}
\toprule
\normalsize
\label{tab:4}
\multirow{2}{*}{\textbf{Model}} & 
\multirow{2}{*}{\textbf{Institute}} & 
\multirow{2}{*}{\textbf{Release}} & 
\multirow{2}{*}{\textbf{Base LLM}} & 
\multirow{2}{*}{\textbf{Base LLM Params}} &        
\multirow{2}{*}{\textbf{Lang.}} &   
\multirow{2}{*}{\textbf{Input Repr.}} & 
\textbf{Pre-train} & 
\textbf{Full-} & 
\multicolumn{2}{c}{\textbf{Multimodality}} \\
\cmidrule(lr){10-11}

& & & & & & & \textbf{Data Scale} & \textbf{Duplex} & Text & Audio \\
\midrule
\YearRow{Year 2022}
dGSLM \cite{nguyen2023generative} & \tablogo{icon/meta.png} & Mar 2022 & - & - & EN & Discrete & 2K Hrs audio & \xmark & \xmark & \cmark \\
\YearRow{Year 2023}
SpeechGPT \cite{zhang2023speechgpt} & \tablogo{icon/fudan.png} & May 2023 & LLaMA-13B & 13B & EN & Discrete & 60K Hrs audio + 9M unit-text pairs + 37,969 quadruplets & \xmark & \cmark & \cmark \\
Pengi \cite{deshmukh2023pengi} & \tablogo{icon/microsoft.png} & May 2023 & GPT-2 & 124M & EN & Contin. & 3.4M audio-text pairs & \xmark & \cmark & \cmark \\
LTU \cite{ramaswamy2025enhancing} & \tablogo{icon/mit.png} & May 2023 & LLaMA-7B & 7B & EN & Contin. & 1.9M closed + 3.7M open-ended AQA pairs & \xmark & \cmark & \cmark \\
Spectron \cite{nachmani2024spoken} & \tablogo{icon/Google.jpg} & May 2023 & - & 350M/1B & EN & Contin. & - & \xmark & \cmark & \cmark \\
AudioPaLM \cite{rubenstein2023audiopalm} & \tablogo{icon/Google.jpg} & Jun 2023 & PaLM-2 & 8B & Multi. & Discrete & - & \xmark & \cmark & \cmark \\
MU-LLaMA \cite{liu2024music} & \tablogo{icon/MULLa.png} & Aug 2023 & LLaMA-2-7B & 7B & EN & Contin. & - & \xmark & \cmark & \cmark \\
LTU-AS \cite{gong2023joint} & \tablogo{icon/mit.png} & Sep 2023 & LLaMA-7B & 7B & EN & Contin. & 9.6M Open-ASQA & \xmark & \cmark & \cmark \\
SLM \cite{wang2023slm} & \tablogo{icon/Google.jpg} & Sep 2023 & mT0-MT XXL & 13B & Multi. & Contin. & - & \xmark & \cmark & \cmark \\
SALMONN \cite{tang2023salmonn} & \tablogo{icon/thu.jpg} & Oct 2023 & Vicuna-13B & 13B & EN, CN & Contin. & 4760 Hrs audio & \xmark & \cmark & \cmark \\
LauraGPT \cite{du2023lauragpt} & \tablogo{icon/alibaba.png} & Oct 2023 & Qwen-1.8B & 2B & EN, CN & Contin. & - & \xmark & \cmark & \cmark \\
Qwen-Audio\cite{chu2023qwen} & \tablogo{icon/Qwen.png} & Nov 2023 & Qwen-7B & 7B & Multi. & Contin. & 130K+ Hrs audio & \xmark & \cmark & \cmark \\
ParalinGPT \cite{lin2024paralinguistics} & \tablogo{icon/alexa.png} & Dec 2023 & DialoGPT & 345M & EN & Contin. & 140 Hrs audio & \xmark & \cmark & \cmark \\
E-chat \cite{xue2024chat} & \tablogo{icon/xbu.png} & Dec 2023 & Baichuan2-7B-Chat & 7B & CN & Contin. & 10K Hrs ASR data & \xmark & \cmark & \cmark \\
\YearRow{Year 2024}
SpeechGPT-Gen \cite{zhang2024speechgpt} & \tablogo{icon/fudan.png} & Jan 2024 & LLaMA-2-7B-Chat & 7B & EN & Discrete & - & \xmark & \cmark & \cmark \\
Audio Flamingo \cite{kong2024audio} & \tablogo{icon/nvidia.png} & Feb 2024 & OPT-IML-1.3B & 1.3B & EN & Contin. & 21K Hrs audio & \xmark & \cmark & \cmark \\
Spoken-LLM \cite{lin2024advancing} & \tablogo{icon/hk.png} & Feb 2024 & Llama-2-7B-Chat & 7B & EN & Contin. & 16,472 current-response speech pairs & \xmark & \cmark & \cmark \\
Spirit LM \cite{nguyen2025spirit} & \tablogo{icon/meta.png} & Feb 2024 & Llama-2-7B & 7B & EN & Discrete & 35.2B tokens & \xmark & \cmark & \cmark \\
USDM \cite{li2024emergent} & \tablogo{icon/seu.jpg} & Feb 2024 & Mistral-7B & 7B & EN & Discrete & 87K Hrs audio & \xmark & \cmark & \cmark \\
WavLLM \cite{hu2024wavllm} & \tablogo{icon/microsoft.png} & Mar 2024 & LLaMA-2-7B-Chat & 7B & EN & Contin. & - & \xmark & \cmark & \cmark \\
SpeechVerse \cite{das2024speechverse} & \tablogo{icon/Amazon.png} & May 2024 & Flan-T5-XL & 3B & EN & Contin. & - & \xmark & \cmark & \cmark \\
GAMA \cite{ghosh2024gama} & \tablogo{icon/mu.png} & Jun 2024 & LLaMA2-7B & 7B & EN & Contin. & 2.2M audio-caption pairs & \xmark & \cmark & \cmark \\
Qwen2-Audio \cite{chu2024qwen2} & \tablogo{icon/Qwen.png} & Jul 2024 & Qwen-7B & 7B & Multi. & Contin. & 520K Hrs audio & \xmark & \cmark & \cmark \\
FunAudioLLM \cite{an2024funaudiollm} & \tablogo{icon/alibaba.png} & Jul 2024 & - & - & Multi. & - & - & \xmark & \cmark & \cmark \\
Mini-Omni \cite{xie2024mini} & \tablogo{icon/thu.jpg} & Aug 2024 & Qwen2-0.5B & 0.5B & - & Discrete & 8K Hrs speech + 2M text examples & \cmark & \cmark & \cmark \\
Moshi \cite{defossez2024moshi} & \tablogo{icon/Moshi.png} & Sep 2024 & Helium & 7B & EN & Discrete & 7M Hrs audio + 2.1T text tokens & \cmark & \cmark & \cmark \\
LLaMA-Omni \cite{fang2024llama} & \tablogo{icon/zky.jpg} & Sep 2024 & Llama-3.1-8B-Instruct & 8B & EN & Contin. & - & \xmark & \cmark & \cmark \\
Parrot \cite{meng2024parrot} & \tablogo{icon/tencent.png} & Sep 2024 & Llama 3.1-8B & 8B & EN & Discrete & 74,554 Hrs audio & \cmark & \xmark & \cmark \\
OmniFlatten \cite{zhang2025omniflatten} & \tablogo{icon/alibaba.png} & Oct 2024 & Qwen2-0.5B & 0.5B & EN, CN & Discrete & - & \cmark & \cmark & \cmark \\
IntrinsicVoice \cite{zhang2024intrinsicvoice} & \tablogo{icon/fudan.png} & Oct 2024 & Qwen2-7B-Instruct & 7B & - & Discrete & 20K Hrs audio & \xmark & \cmark & \cmark \\
DiVA \cite{held2025distilling} & \tablogo{icon/GT.jpg} & Oct 2024 & Llama 3 & 8B & EN & Contin. & - & \xmark & \cmark & \cmark \\
Freeze-Omni \cite{wang2024freeze} & \tablogo{icon/vita.png} & Nov 2024 & Qwen2-7B-Instruct & 7B & EN, CN & Contin. & - & \cmark & \cmark & \cmark \\
GLM-4-Voice \cite{zeng2024glm} & \tablogo{icon/GLM.png} & Dec 2024 & GLM-4-9B & 9B & EN, CN & Discrete & 1T tokens & \xmark & \cmark & \cmark \\
KE-Omni \cite{KeOmniR2025} & \tablogo{icon/ke.png} & Dec 2024 & LLaMA-3.1-8B-Instruct & 8B & EN, CN & Contin. & - & \xmark & \cmark & \cmark \\
MERaLiON-Audio \cite{he2024meralion} & \tablogo{icon/mer.png} & Dec 2024 & SEA-LION V3 & 10B & Multi. & Contin. & - & \xmark & \cmark & \cmark \\
\bottomrule
\end{tabular}
}
\end{table}

\clearpage

\renewcommand{\TableCols}{11}
\begin{table}[!t]
\addtocounter{table}{-1}
\caption[]{Summary of Large Audio Language Models from 2022 to 2026 (continued)}
\centering
\normalsize
\renewcommand{\arraystretch}{1.6}
\resizebox{\textwidth}{!}{
\begin{tabular}{l c c l c c c c c c c}
\toprule
\multirow{2}{*}{\textbf{Model}} &
\multirow{2}{*}{\textbf{Institute}} &
\multirow{2}{*}{\textbf{Release}} &
\multirow{2}{*}{\textbf{Base LLM}} &
\multirow{2}{*}{\textbf{Base LLM Params}} &
\multirow{2}{*}{\textbf{Lang.}} &
\multirow{2}{*}{\textbf{Input Repr.}} &
\textbf{Pre-train} &
\textbf{Full-} &
\multicolumn{2}{c}{\textbf{Multimodality}} \\
\cmidrule(lr){10-11}

& & & & & & & \textbf{Data Scale} & \textbf{Duplex} & Text & Audio \\
\midrule
\YearRow{Year 2025}
MinMo \cite{zhang2025mimo} & \tablogo{icon/Qwen.png} & Jan 2025 & Qwen2.5-7B-Instruct & 7B & Multi. & Contin. & - & \cmark & \cmark & \cmark \\
FireRedASR \cite{shi2026qwen3} & \tablogo{icon/xiaohongshuLOGO.png} & Jan 2025 & Qwen2-7B-Instruct & 7B & Multi. & Contin. & - & \xmark & \cmark & \cmark \\
Step-Audio \cite{tian2025step} & \tablogo{icon/step.png} & Feb 2025 & Step-1 & 130B & Multi. & Discrete & 3.3T tokens & \xmark & \cmark & \cmark \\
Baichuan-Audio \cite{li2025baichuan} & \tablogo{icon/baichuan.jpg} & Feb 2025 & Baichuan-Audio-Base & 7B & EN, CN & Discrete & 887K Hrs audio + 100B tokens & \xmark & \cmark & \cmark \\
Audio Flamingo 2 \cite{ghosh2025audio} & \tablogo{icon/nvidia.png} & Mar 2025 & Qwen2.5-3B & 3B & EN & Contin. & 8M+ audio-caption pairs & \xmark & \cmark & \cmark \\
Kimi-Audio \cite{ding2025kimi} & \tablogo{icon/kimi.png} & Apr 2025 & Qwen2.5-7B & 7B & EN, CN & Hybrid & 13M+ Hrs audio & \xmark & \cmark & \cmark \\
VITA-Audio \cite{long2025vita} & \tablogo{icon/vita.png} & May 2025 & Qwen2.5-7B-Instruct & 7B & EN, CN & Discrete & 200K Hrs audio & \xmark & \cmark & \cmark \\
Step-Audio 2 \cite{wu2025step} & \tablogo{icon/step.png} & Jul 2025 & - & - & Multi. & Contin. & 680B tokens and 8M Hrs audio & \xmark & \cmark & \cmark \\
Audio Flamingo 3 \cite{goel2025audio} & \tablogo{icon/nvidia.png} & Jul 2025 & Qwen2.5-7B & 7B & EN & Contin. & - & \xmark & \cmark & \cmark \\
DeSTA2.5-Audio \cite{lu2026desta2} & \tablogo{icon/hk.png} & Jul 2025 & Llama3.1-8B-Instruct & 8B & EN & Contin. & 7K Hrs audio & \xmark & \cmark & \cmark \\
FireRedChat \cite{chen2025fireredchat} & \tablogo{icon/xiaohongshuLOGO.png} & Sep 2025 & Qwen2.5 & - & EN, CN & - & - & \cmark & \cmark & \cmark \\
Falcon3-Audio \cite{kumar2025competitive} & \tablogo{icon/tii.png} & Sep 2025 & Falcon3-Instruct & 1/3/7B & EN & Contin. & - & \xmark & \cmark & \cmark \\
Step-Audio-R1 \cite{tian2025step} & \tablogo{icon/step.png} & Nov 2025 & Qwen2.5-32B & 32B & EN, CN & Contin. & 1.356T tokens & \xmark & \cmark & \cmark \\
Step-Audio-EditX \cite{yan2025step} & \tablogo{icon/step.png} & Nov 2025 & - & 3B & Multi. & Discrete & - & \xmark & \cmark & \cmark \\
SeaLLMs-Audio \cite{liu2025seallms} & \tablogo{icon/SeaLLM-Audio.png} & Nov 2025 & Qwen2.5-7B & 7B & Multi. & Contin. & - & \xmark & \cmark & \cmark \\
Fun-Audio-Chat \cite{team2025fun} & \tablogo{icon/Qwen.png} & Dec 2025 & Qwen3 & 8/30B & EN, CN & Discrete & - & \xmark & \cmark & \cmark \\
MiMo-Audio \cite{zhang2025mimo} & \tablogo{icon/Mi.png} & Dec 2025 & MiMo-7B-Base & 7B & Multi. & Discrete & 100M+ Hrs audio & \xmark & \cmark & \cmark \\
\YearRow{Year 2026}
Step-Audio-R1.1 \cite{tian2025step} & \tablogo{icon/step.png} & Jan 2026 & Qwen2.5-32B & 32B & EN, CN & Contin. & 1.356T tokens & \xmark & \cmark & \cmark \\
Qwen3-ASR \cite{shi2026qwen3} & \tablogo{icon/Qwen.png} & Jan 2026 & Qwen3 & 0.6/1.7B & Multi. & Contin. & 40M Hrs audio + 3T tokens & \xmark & \cmark & \cmark \\
Covo-Audio \cite{wang2026covo} & \tablogo{icon/tencent.png} & Feb 2026 & Qwen2.5-7B-Base & 7B & EN, CN & Contin. & 2T tokens & \cmark & \cmark & \cmark \\
Eureka-Audio \cite{zhang2026eureka} & \tablogo{icon/baidu.png} & Feb 2026 & Qwen3-1.7B-Base & 1.7B & EN, CN & Contin. & 1T tokens & \xmark & \cmark & \cmark \\
Audio Flamingo Next \cite{ghosh2026audio} & \tablogo{icon/nvidia.png} & Apr 2026 & Qwen2.5-7B & 7B & Multi. & Contin. & 363.5K Hrs audio & \xmark & \cmark & \cmark \\
Step-Audio-R1.5 \cite{zhang2026step} & \tablogo{icon/step.png} & Apr 2026 & Qwen2.5-32B & 32B & EN, CN & Contin. & - & \xmark & \cmark & \cmark \\
DuplexSLA \cite{zhang2026duplexsla} & \tablogo{icon/step.png} & May 2026 & Qwen2.5-7B & 7B & EN, CN & Contin. & 500K Hrs audio + 1.92M text samples & \cmark & \cmark & \cmark \\
StepAudio 2.5 \cite{lin2026stepaudio} & \tablogo{icon/step.png} & May 2026 & - & - & EN, CN & Contin. & 2.2T text and audio tokens & \xmark & \cmark & \cmark \\
MOSS-Audio \cite{yang2026moss} & \tablogo{icon/sii.png} & Jun 2026 & Qwen3-4B/8B & 4B/8B & EN, CN & Contin. & 1.2T tokens & \xmark & \cmark & \cmark \\
BayLing-Duplex \cite{fang2026bayling} & \tablogo{icon/ict.png} & Jun 2026 & GLM-4-9B & 9B & EN, CN & Discrete & - & \cmark & \cmark & \cmark \\
FlexiSLM \cite{li2026flexislm} & \tablogo{icon/cuhksz.png} & Jun 2026 & Qwen2.5-7B-Instruct & 7B & EN & Contin. & - & \xmark & \cmark & \cmark \\

\bottomrule
\end{tabular}
}
\vspace{1mm}
\begin{flushleft}
\footnotesize
\textit{Note:} This table only summarizes large language models with audio modality, excluding models with image or video modality support. 
``Lang.'' is short for language, where ``EN'' denotes English, ``CN'' denotes Chinese, and ``Multi.'' is short for multiple languages, indicating support for more than two languages. 
``Input Repr.'' is short for input representation, and ``Contin.'' is short for continuous representation.
\end{flushleft}
\end{table}

\subsection{Detailed Comparison of LALM Evaluation Benchmarks}
\label{app:detailed-benchmarks}

Table~\ref{tab:5} extends the benchmark coverage summary in
Table~\ref{tab:3} by reporting the original evaluation metrics
used by each benchmark. The general-capability dimensions comprise perception,
reasoning, and interaction, while the trustworthiness dimensions comprise
hallucination, privacy, authentication, safety, robustness, and fairness.

A check mark is assigned only when a benchmark explicitly includes evaluation
examples, annotations, or metrics for the corresponding dimension. A dimension
that is mentioned only as motivation or future work is not marked as covered.
This criterion is applied consistently to both the compact and detailed tables.

\renewcommand{\TableCols}{12}
\begin{table}[!t]
\centering
\normalsize
\setlength{\tabcolsep}{4pt}
\renewcommand{\arraystretch}{1.6}
\caption{Overview of LALM evaluation benchmarks across general capabilities and trustworthy dimensions.}
\label{tab:5}
\begin{adjustbox}{max width=\textwidth}
\begin{tabular}{@{}lclccccccccc@{}}
\toprule

\multirow{2}{*}{\textbf{Benchmark}} &
\multirow{2}{*}{\textbf{Release}} &
\multirow{2}{*}{\textbf{Metrics}} &
\multicolumn{3}{c}{\textbf{General}$^\dagger$} &
\multicolumn{6}{c}{\textbf{Trustworthy}$^\ddagger$} \\
\cmidrule(lr){4-6} \cmidrule(lr){7-12}

& & &
\textbf{PE} & \textbf{RE} & \textbf{IN} &
\textbf{H} & \textbf{P} & \textbf{A} & \textbf{S} & \textbf{R} & \textbf{F} \\
\midrule
\YearRow{Year 2024}
AudioBench \cite{wang2025audiobench}        & Jan 2024 & WER, METEOR, LLM-as-a-Judge & \cmark & \cmark & \xmark & \xmark & \xmark & \xmark & \xmark & \cmark & \xmark \\
MMAU \cite{sakshi2024mmau}                  & Oct 2024 & Accuracy & \cmark & \cmark & \xmark & \xmark & \xmark & \xmark & \xmark & \xmark & \xmark \\
VoiceBench \cite{chen2026voicebench}             & Oct 2024 & LLM-as-a-Judge, Accuracy, Refusal Rate & \xmark & \cmark & \xmark & \xmark & \xmark & \xmark & \xmark & \cmark & \xmark \\
\midrule
\YearRow{Year 2025}
Jailbreak-AudioBench \cite{cheng2025jailbreak}  & Jan 2025 & ASR & \xmark & \xmark & \xmark & \xmark & \xmark & \xmark & \cmark & \xmark & \xmark \\
URO-Bench \cite{yan2025uro}             & Feb 2025 & LLM-as-a-Judge, UTMOS, WER/CER, First Packet Latency & \cmark & \cmark & \cmark & \xmark & \cmark & \cmark & \cmark & \xmark & \xmark \\
S2S-Arena \cite{jiang2025s2s}                     & Mar 2025 & ELO Rating, Win Rate, Inter-Annotator Agreement & \cmark & \xmark & \cmark & \xmark & \xmark & \xmark & \xmark & \xmark & \xmark \\
Talking Turns \cite{arora2025talking}             & Mar 2025 & Agreement, Accuracy, ROC-AUC, F1 (judge model) & \cmark & \cmark & \cmark & \xmark & \xmark & \xmark & \xmark & \xmark & \xmark \\
MMAR \cite{ma2025mmar}                  & May 2025 & Accuracy & \cmark & \cmark & \xmark & \xmark & \xmark & \xmark & \xmark & \cmark & \xmark \\
SAKURA \cite{yang2025sakura}                  & May 2025 & Accuracy, LLM-as-a-Judge & \cmark & \cmark & \xmark & \xmark & \xmark & \xmark & \xmark & \xmark & \xmark \\
VocalBench \cite{liu2025vocalbench}             & May 2025 & Accuracy, LLM-as-a-Judge, UTMOS, WER, Refusal Rate, Preserve Rate & \cmark & \cmark & \cmark & \xmark & \xmark & \xmark & \cmark & \cmark & \xmark \\
JALMBench \cite{peng2025jalmbench}             & May 2025 & ASR, Attack Efficiency & \xmark & \xmark & \xmark & \xmark & \xmark & \xmark & \cmark & \xmark & \xmark \\
AHaBench \cite{cheng2026aha}             & May 2025 & AHa-Score, m-AHa, GPT-4-Judge Consistency, POPE & \xmark & \xmark & \cmark & \cmark & \xmark & \xmark & \xmark & \xmark & \xmark \\
AudioJailbreak \cite{chen2025audiojailbreak}  & May 2025 & ASR & \xmark & \xmark & \xmark & \xmark & \xmark & \xmark & \cmark & \xmark & \xmark \\
AJailBench \cite{song2025audio}            & May 2025 & ASR, TS, PV, Relevance, Similarity & \xmark & \xmark & \xmark & \xmark & \xmark & \xmark & \cmark & \xmark & \xmark \\
VocalAgent \cite{kim2025vocalagent}            & May 2025 & macro-F1, Accuracy, FPR, Refusal Rate, Goodness@0.1 & \cmark & \xmark & \xmark & \xmark & \xmark & \xmark & \xmark & \cmark & \xmark \\
AudioTrust \cite{li2025audiotrust}            & May 2025 & GPT-4o Score, CM-WER, CCR, DSR, HRR, FAR, SES, Refusal Rate, Group Unfairness Score & \xmark & \xmark & \xmark & \cmark & \cmark & \cmark & \cmark & \cmark & \cmark \\
MMSU \cite{wang2025mmsu}      & Jun 2025 & Accuracy & \cmark & \cmark & \xmark & \xmark & \xmark & \xmark & \xmark & \xmark & \xmark \\
SOVA-Bench \cite{hou2025sova}                    & Jun 2025 & Accuracy, WER, GPTEval, LLM-as-a-Judge, UTMOSv2 & \cmark & \cmark & \cmark & \xmark & \xmark & \xmark & \xmark & \xmark & \xmark \\
WildSpeech-Bench \cite{zhang2025wildspeechbench}       & Jun 2025 & LLM-as-a-Judge, UTMOS, Query-Aware Checklist & \cmark & \cmark & \cmark & \xmark & \xmark & \xmark & \xmark & \cmark & \xmark \\
ContextASR-Bench \cite{wang2025contextasr}      & Jul 2025 & WER, NE-WER, NE-FNR & \cmark & \xmark & \xmark & \xmark & \xmark & \xmark & \xmark & \xmark & \xmark \\
C\textsuperscript{3}\cite{ma2025c3}      & Jul 2025 & LLM-as-a-Judge, Human Evaluation Score & \cmark & \cmark & \cmark & \xmark & \xmark & \xmark & \xmark & \cmark & \xmark \\
WoW-Bench \cite{kim2025wow}             & Aug 2025 & Accuracy & \cmark & \cmark & \xmark & \xmark & \xmark & \xmark & \xmark & \xmark & \xmark \\
MMAU-Pro \cite{kumar2025mmau}              & Aug 2025 & Accuracy & \cmark & \cmark & \cmark & \xmark & \xmark & \xmark & \xmark & \xmark & \xmark \\
MCR-BENCH \cite{wang2025audio}             & Aug 2025 & Accuracy, Norm Acc, Macro Acc, TIR, MRS & \cmark & \cmark & \xmark & \cmark & \xmark & \xmark & \xmark & \xmark & \xmark \\
SpeechR \cite{yang2025speechr}               & Aug 2025 & Accuracy, Logical Consistency Score & \xmark & \cmark & \xmark & \xmark & \xmark & \xmark & \xmark & \xmark & \xmark \\
AudioSafe \cite{lin2025hidden}   & Aug 2025 & Accuracy, ASR & \xmark & \xmark & \xmark & \xmark & \xmark & \xmark & \cmark & \cmark & \xmark \\
VoiceAssistant-Eval \cite{wang2025voiceassistant} & Sep 2025 & LLM-as-a-Judge, UTMOS, WER, Speaker Similarity & \cmark & \cmark & \cmark & \xmark & \xmark & \xmark & \cmark & \cmark & \xmark \\
Gaslighting Attacks \cite{wu2026benchmarking} & Sep 2025 & Accuracy, Behavioral Consistency & \xmark & \xmark & \cmark & \xmark & \xmark & \xmark & \xmark & \cmark & \xmark \\
MUSE \cite{carone2025muse}                  & Oct 2025 & Accuracy & \cmark & \cmark & \xmark & \xmark & \xmark & \xmark & \xmark & \xmark & \xmark \\
LISTEN \cite{chen2025audio}                & Oct 2025 & Accuracy & \cmark & \cmark & \xmark & \cmark & \xmark & \xmark & \xmark & \xmark & \xmark \\
AudioMarathon \cite{he2025audiomarathon}         & Oct 2025 & F1, WAR & \cmark & \cmark & \xmark & \xmark & \xmark & \xmark & \xmark & \xmark & \xmark \\
ISA-Bench \cite{li2025isa}             & Oct 2025 & IFR, WER, BLEU, ACC, METEOR, RPS & \cmark & \xmark & \xmark & \xmark & \xmark & \xmark & \cmark & \xmark & \xmark \\
Hearing the Order \cite{lin2025hearing}     & Oct 2025 & Accuracy, CKLD & \xmark & \xmark & \xmark & \xmark & \xmark & \xmark & \xmark & \cmark & \xmark \\
Safety under Emotional Variations \cite{feng2025investigating}
                      & Oct 2025 & NRR, Unsafe Rate & \xmark & \xmark & \xmark & \xmark & \xmark & \xmark & \cmark & \cmark & \xmark \\
Gender Bias in SpeechLLMs \cite{satish2025bias}
                      & Oct 2025 & MCQA Selection Rate, LLM-as-a-Judge & \xmark & \xmark & \xmark & \xmark & \xmark & \xmark & \xmark & \xmark & \cmark \\
BRACE \cite{guo2025brace}                & Dec 2025 & F1 & \cmark & \xmark & \xmark & \cmark & \xmark & \xmark & \xmark & \xmark & \xmark \\
Spoken DialogSum \cite{lu2025spoken}      & Dec 2025 & ROUGE-L & \xmark & \xmark & \cmark & \xmark & \xmark & \xmark & \xmark & \xmark & \xmark \\
MAC-SLU \cite{peng2025mac}               & Dec 2025 & Accuracy, F1 & \cmark & \xmark & \xmark & \xmark & \xmark & \xmark & \xmark & \xmark & \xmark \\
SH-Bench \cite{zhan2025protecting}      & Dec 2025 & SE (Selective Efficacy), Accuracy & \cmark & \xmark & \xmark & \xmark & \cmark & \xmark & \xmark & \xmark & \xmark \\
\bottomrule
\end{tabular}
\end{adjustbox}
\end{table}

\clearpage

\renewcommand{\TableCols}{12}
\begin{table}[!t]
\centering
\normalsize
\setlength{\tabcolsep}{4pt}
\renewcommand{\arraystretch}{1.6}
\addtocounter{table}{-1}
\caption[]{Overview of LALM evaluation benchmarks across general capabilities and trustworthy dimensions (continued).}
\begin{adjustbox}{max width=\textwidth}
\begin{tabular}{@{}lclccccccccc@{}}
\toprule

\multirow{2}{*}{\textbf{Benchmark}} &
\multirow{2}{*}{\textbf{Release}} &
\multirow{2}{*}{\textbf{Metrics}} &
\multicolumn{3}{c}{\textbf{General}$^\dagger$} &
\multicolumn{6}{c}{\textbf{Trustworthy}$^\ddagger$} \\
\cmidrule(lr){4-6} \cmidrule(lr){7-12}

& & &
\textbf{PE} & \textbf{RE} & \textbf{IN} &
\textbf{H} & \textbf{P} & \textbf{A} & \textbf{S} & \textbf{R} & \textbf{F} \\
\midrule
\YearRow{Year 2026}
WESR \cite{yang2026wesr}                  & Jan 2026 & Precision, Recall, F1, WER & \cmark & \xmark & \xmark & \xmark & \xmark & \xmark & \xmark & \xmark & \xmark \\
RSA-Bench \cite{zhang2026rsa}             & Jan 2026 & WER, Accuracy, LLM-as-a-Judge & \cmark & \cmark & \xmark & \xmark & \xmark & \xmark & \xmark & \cmark & \xmark \\
PALM-Bench \cite{wang2026palm}            & Jan 2026 & BLEU, F1-score, BERTScore, LLM-as-a-Judge & \xmark & \cmark & \cmark & \cmark & \xmark & \xmark & \xmark & \xmark & \xmark \\
ChronosAudio \cite{luo2026chronosaudio}          & Jan 2026 & WER, BERTScore, Acc\_dict, LS, TS, MS, LLM-as-a-Judge & \cmark & \cmark & \xmark & \cmark & \xmark & \xmark & \xmark & \xmark & \xmark \\
HearSay \cite{wang2026hearsay}               & Jan 2026 & IAR, ARR, BBR & \xmark & \xmark & \xmark & \xmark & \cmark & \xmark & \xmark & \xmark & \xmark \\
AGL1K \cite{zhang2026sonar}      & Jan 2026 & Accuracy, Geoscore, Mean Distance Error, Thresholded Accuracy, Reject Rate & \xmark & \cmark & \xmark & \xmark & \xmark & \xmark & \xmark & \xmark & \cmark \\
VoxPrivacy \cite{wang2026voxprivacy}      & Jan 2026 & Accuracy, Precision, Recall, F1-Score, IRR, EER & \xmark & \xmark & \cmark & \xmark & \cmark & \cmark & \xmark & \xmark & \xmark \\
BiasInEar \cite{wei2026bias} & Feb 2026 & Accuracy, Entropy, APES, Fleiss' $\kappa$ & \xmark & \xmark & \xmark & \xmark & \xmark & \xmark & \xmark & \xmark & \cmark \\
DailyTalkEdit \cite{xu2026holiantispoof}           & Feb 2026 & Accuracy, EER, F1, Seg-F1, LLM-as-a-Judge & \xmark & \xmark & \xmark & \xmark & \xmark & \cmark & \xmark & \xmark & \xmark \\
HumDial-EIBench \cite{wang2026humdialeibenchhumanrecordedmultiturnemotional}               & Apr 2026 & Accuracy, LLM-as-Judge & \xmark & \cmark & \cmark & \xmark & \xmark & \xmark & \xmark & \cmark & \xmark \\
VoxSafeBench \cite{wang2026voxsafebench}               & Apr 2026 & Safety Awareness Score, JSR, Perception Probing Accuracy & \cmark & \xmark & \cmark & \xmark & \cmark & \xmark & \cmark & \xmark & \cmark \\
HalluAudio \cite{zhao2026halluaudiocomprehensivebenchmarkhallucination} 
& Apr 2026 
& Accuracy, Hallucination Rate, Yes/No Bias, Error-type Analysis, FRR 
& \cmark & \cmark & \xmark 
& \cmark & \xmark & \xmark & \xmark & \cmark & \xmark \\

MUSA \cite{koo2026largeaudiolanguagemodels} & May 2026 & Accuracy, ECE
& \cmark & \cmark & \xmark
& \xmark & \xmark & \xmark & \xmark & \cmark & \xmark \\
PitchBench \cite{dujardin2026pitchbenchmeasuringpitchhearing} & May 2026 & Accuracy
& \cmark & \xmark & \xmark
& \xmark & \xmark & \xmark & \xmark & \cmark & \xmark \\
EnvMem \cite{xiao2026cantrememberuncoveringrepresentation} & May 2026 & Accuracy, $\Delta(N)$, CKA, $\mathrm{cov}_{90}$
& \cmark & \xmark & \cmark
& \xmark & \xmark & \xmark & \xmark & \cmark & \xmark \\

VoiceGiraffe \cite{ye2026voicegiraffebenchmarkextremelongcontext} & May 2026 & Macro-Average Accuracy
& \cmark & \cmark & \xmark
& \xmark & \xmark & \xmark & \xmark & \cmark & \xmark \\

VoxParadox \cite{pang2026voxparadox} & May 2026 & GT Accuracy, ALA
& \cmark & \xmark & \xmark
& \cmark & \xmark & \xmark & \xmark & \cmark & \xmark \\

MusTBENCH \cite{kwon2026mustbenchbenchmarkingadvancingtemporal} & May 2026 & Accuracy, Hit@$T$, METEOR, CLAP Score, Temporal IoU, F1
& \cmark & \cmark & \xmark
& \cmark & \xmark & \xmark & \xmark & \xmark & \xmark \\

SpeechJBB \cite{ceccatelli2026speechjbbprobingsafetyalignment} & Jun 2026 & RR, DR, JSR
& \cmark & \xmark & \xmark
& \xmark & \xmark & \xmark & \cmark & \cmark & \xmark \\

GlobeAudio \cite{tan2026globeaudiomultilingualmulticulturalbenchmark} & Jun 2026 & Accuracy
& \cmark & \cmark & \xmark
& \xmark & \xmark & \xmark & \xmark & \cmark & \cmark \\

AudioProcessBench \cite{zhao2026audioprocessbenchbenchmarkidentifyingprocess} & Jun 2026 & PRMScore, Balanced Accuracy, AUROC, BoN/MV Accuracy
& \cmark & \cmark & \xmark
& \cmark & \xmark & \xmark & \xmark & \xmark & \xmark \\

RAIL \cite{jin2026railrethinkingauditoryintelligence} & Jun 2026 & Accuracy, LLM-as-a-Judge, B-AUC
& \cmark & \cmark & \xmark
& \xmark & \xmark & \xmark & \xmark & \xmark & \xmark \\

IHBench \cite{salimi2026ihbenchevaluatingpostinterruptionrecovery} & Jun 2026 & TF Win Rate, RQ Pass Rate
& \xmark & \cmark & \cmark
& \xmark & \xmark & \xmark & \xmark & \cmark & \xmark \\

AOR-Bench \cite{yang2026aorbenchlargeaudiolanguage} & Jun 2026 & ORR, TRR, MB-Score
& \cmark & \cmark & \xmark
& \xmark & \xmark & \xmark & \cmark & \cmark & \xmark \\

MSU-Bench \cite{sun2026msubenchspeakercentricunderstandingconversational} & Jun 2026 & Exact-Match Accuracy, Error-type Analysis
& \cmark & \cmark & \cmark
& \cmark & \xmark & \xmark & \xmark & \xmark & \xmark \\

ParaPairAudioBench \cite{jeon2026parapairaudiobenchparalinguisticpairwiseaudio} & Jun 2026 & Accuracy, Acc@A, Acc@B, Position-Bias Gap
& \cmark & \xmark & \xmark
& \xmark & \xmark & \xmark & \xmark & \cmark & \xmark \\

CASU \cite{zhang2026soundsscenesbenchmarkevaluating} & Jun 2026 & BLEU-4, BERTScore, LLM-as-a-Judge, 1-WER, Accuracy
& \cmark & \cmark & \xmark
& \cmark & \xmark & \xmark & \xmark & \xmark & \xmark \\

SpeechEQ \cite{wu2026speecheqbenchmarkingemotionalintelligence} & Jun 2026 & Acc$_1$, Acc$_2$, Acc$_{\mathrm{traj}}$, SEQ
& \cmark & \cmark & \cmark
& \xmark & \xmark & \xmark & \xmark & \cmark & \xmark \\

\bottomrule
\end{tabular}
\end{adjustbox}
\vspace{0.8ex}
\parbox{\textwidth}{%
\scriptsize
$^\dagger$ \textbf{General}: Perception (PE), Reasoning (RE), Interaction (IN). \\
$^\ddagger$ \textbf{Trustworthy}: Hallucination (H), Privacy (P), Authentication (A), Safety (S), Robustness (R), Fairness (F).
}
\end{table}

\end{document}